\documentclass[12pt]{article}
\usepackage[top=0.7in,bottom=0.7in,left=0.7in,right=0.7in]{geometry}
\usepackage{color}
\usepackage{indentfirst}
\usepackage{latexsym,bm} 
\usepackage{amsmath,amssymb} 
\usepackage{graphicx}
\usepackage{multirow}
\usepackage[justification=centering]{caption} 
\usepackage{textcomp} 
\usepackage{stmaryrd} 
\usepackage{caption}
\usepackage{hyperref}
\usepackage[numbers]{natbib}
\hypersetup{
	colorlinks=true
	linktoc=all
	}
\usepackage[dvipsnames]{xcolor}

\begin{document}

\title{Unbounded entropy production and violent fragmentation for repulsive-to-attractive interaction quench in long-range interacting systems}

\author{Paolo Molignini$^1$~\footnote{Corresponding author, E-mail: \url{paolo.molignini@fysik.su.se}}, Barnali Chakrabarti$^2$ \\
\textit{ {\tiny $^1$ Department of Physics, Stockholm University, AlbaNova University Center, 106 91 Stockholm, Sweden}} \\
\textit{ {\tiny $^2$Department of Physics, Presidency University, 86/1   College Street, Kolkata 700073, India}}
}

\date{}
\maketitle

\begin{abstract}
We study the non-equilibrium dynamics of a one-dimensional Bose gas with long-range interactions that decay as $(\frac{1}{r^{\alpha}})$ $(0.5 < \alpha <4.0$). 
We investigate exotic dynamics when the interactions are suddenly switched from strongly repulsive to strongly attractive, a procedure known to generate super-Tonks-Girardeau gases in systems with contact interactions.
We find that relaxation is achieved through a complex intermediate dynamics demonstrated by violent fragmentation and chaotic delocalization. 
We establish that the relaxed state exhibits classical gaseous characteristics and an asymptotic state associated with unbounded entropy production. 
The phase diagram shows an exponential boundary between the coherent (quantum) gas and the chaotic (classical) gas. 
We show the universality of the dynamics by also presenting analogous results for spinless fermions. 
Weaker quench protocols give a certain degree of control over the relaxation process and induce a slower initial entropy growth.
Our study showcases the complex relaxation behavior of tunable long-range interacting systems that could be engineered in state-of-the-art experiments, e.g. in trapped ions or Rydberg atoms. 
\end{abstract}

\newpage

\newpage 
\section{Introduction} 
\label{sec:intro}

Quantum systems with long-range interactions that decay as $\frac{1}{r^{\alpha}}$ have recently received increasing interest due to rapid developments of experimental techniques for controlling and manipulating atomic, molecular and optical (AMO) systems~\cite{Defenu2021}.
Long-range quantum systems are currently being realized in several experimental platforms such as Rydberg atoms~\cite{Saffman}, trapped ions~\cite{Monroe:2021,Schneider:2012}, polar molecules~\cite{Carr:2009}, and dipolar quantum gases~\cite{Lahaye:2009}.
Furthermore, long-range interactions with a tunable exponent $\alpha$ can be realized in Paul traps~\cite{Islam:2013,Jurcevic:2014}. 
Out of all these platforms, trapped ions present the most unique possibility of exploring long-range interactions with a decay constant $\alpha$ tunable between 0 and 3.  
Moreover, recent experimental setups with ion traps enable the realization of spin chains with controlled long-range interactions~\cite{Cirac,Britton,Islam}. 

The dynamical properties of long-range interacting systems significantly differ from those of short-range interacting ones in many aspects~\cite{Jens:2013,Schachenmayer:2013,Santos:2016,Anton:2016}.
Depending on the exponent $\alpha$, long-range interactions can exhibit novel features.
This is especially true in the very long-ranged regime when $\alpha < d$ ($d$ being the dimension of the system). 
These include long-lived metastable states~\cite{Stefan,Michael}, supersolids and dipolar supersolids~\cite{Li:2017, Boettcher:2019, Tanzi:2019, Tanzi:2019-2, Chomaz:2019, Natale:2019, Tanzi:2021, Norcia:2021, Sohmen:2021,  Sanchez-Baena:2023, Recati:2023},  dynamical phenomena such as time crystals~\cite{Kessler:2021, Kongkhambut:2022,Choi,Zhang} and various Floquet phases~\cite{Li:2019, Wintersperger:2020, Bracamontes:2022, Sun:2023, Zhang:2023}. The nonequilibrium dynamics and relaxation process of isolated quantum systems with short-range interactions is rather well understood and has been extensively studied~\cite{Eriko, Takuya, Kaminishi,Langen:2016,Tang:2018,Marcos:2019,Kaminishi}. 
On the other hand, long-range interactions introduce more intriguing complexities which are poorly understood~\cite{Worm,Kastner, Mori, Defenu2021}. 
For example, short-range interacting systems typically give rise to a single time scale of relaxation, while for strong long-range interactions relaxation can occur in multiple time steps~\cite{Mori}. 

The simplest protocol for probing exotic nonequilibrium dynamics is a quantum quench, which consists of preparing the initial setup in the ground state of a given Hamiltonian and then suddenly switch one of its parameters to a different value at time zero.
A quantum quench allows to probe the fate of a quantum system when perturbed away from equilibrium.
Thermalization~\cite{Rigol} happens when a local observable approaches a statistical value or equilibrium~\cite{Peter,Anthony}. 
However, depending on the quench parameter, an initial quantum system can exhibit a different relaxation process instead.
One example is a phenomenon termed prethermalization~\cite{Berges}, where the local observable first settles to a quasi-stationary value, and only at a later stage reaches thermalization.   
Prethermalization happens when there is a clear separation in relevant time scales with different physical origins, which typically depend on individual systems~\cite{Mori:2018, Gring:2012}. 

Exact theoretical studies on the relaxation pathway in long-range interacting systems for both bosons and fermions are very limited~\cite{Francisco,Schachenmayer:2013}. 
Studies on the quench dynamics with long-range interacting spin chains so far concluded that prethermalization occurs for $0<\alpha<1$~\cite{Mori}.
In another theoretical study, utilizing spatial inhomogeneity, three different relaxation processes were observed: prethermalization only, prethermalization followed by thermalization, and thermalization only~\cite{Gong:2013}.
Prethermalization has also captured the attention in spinless fermions and in the context of fermionic Hubbard models~\cite{Marcus,Martin,Bruno}. 
Although significant amount of work both theoretical and experimental has been devoted to understanding the prethermalization process, its physical origin is still elusive. 
It is also not clear whether it can be related to the quasi-integrability of the system. 
For bosons, theoretical studies so far addressed the cases of quasi-BEC regimes~\cite{Mori:2018} and the case of strongly interacting dipolar bosons remains rather unexplored.
 
Our main motivation is to establish an all-in-one out-of-equilibrium dynamics for the generalized long-range interactions that can be observed in AMO experiments with magnetic atoms~\cite{Chomaz:2023,Patscheider} and polar molecules~\cite{Kaufman:2023,Li-JR:2023}.
To that end, the present work focuses on generalized long-range interacting bosons and fermions where the interactions are quenched from the strongly repulsive to the strongly attractive regime.
This procedure is akin to the approaches used to study the stability of strongly interacting gases with contact interactions starting from the Tonks-Girardeau (TG) regime. 
The strongly interacting bosons in TG limit exhibits remarkable effect of quantum correlation~\cite{Lieb}. 
Due to strong repulsive interactions, atoms escape their spatial overlap to minimize their interaction energy and thereby acquire fermionic properties~\cite{Lieb}, which have been experimentally observed~\cite{Kinoshita:2004}. By quenching the interactions to the strongly \emph{attractive} regime instead makes a transition to another highly excited and correlated state, called a super-Tonks-Girardeau gas (sTG)~\cite{Astrakharchik:2005,Astrakharchik:2008} which has been verified experimentally in the work of Haller {\it et al.}~\cite{Haller:2009}.
The nature of this highly correlated gas can be characterized by either breathing dynamics or confirmation that it neither clusters, nor remains bound, but it is metastable~\cite{Muth:2010, Tschischik:2015}. 
Theoretically, quench problems like the ones described above have been addressed by solving the dynamics of a Lieb-Liniger Bose gas~\cite{Chen:2010}, but several paradigmatic questions regarding relaxation dynamics are still not clarified.

In this work, we follow the experimental protocol of Haller~\cite{Haller:2009} and consider generic power-law interactions with $0.5 \leq \alpha < 4$. For each value of $\alpha$, we first prepare $N=5$ strongly repulsive bosons in the ground state of a harmonic trap, obtaining a crystal state.
Then, at time $t$=0, we suddenly quench their interactions to the strongly attractive regime and monitor the out-of-equlibrium dynamics. 
We solve the time-dependent Schr\"odinger equation (TDSE) for the quench process numerically by employing the MultiConfiguration Time-Dependent Hartree method for bosons (MCTDHB)~\cite{Streltsov:2006,Streltsov:2007,Alon:2007,Alon:2008}. 
The many-body dynamics and the relaxation process are further studied by the evolution of several measures of many-body information entropy which answer the following questions:
i) What is the underlying microscopic dynamics that transforms the strongly repulsive phase into the strongly attractive one?
ii) How does the non equilibrium dynamics change when longer-ranged interactions are employed?
iii) How can the excited gas phase be characterized from a many-body perspective?
iv) Does equilibration happen in a single or dual time scale? 
v) Is it possible to control the relaxation process in the quench dynamics?
Tuning the long-range interactions~\cite{Lin:2023}, we address all these questions by studying quenches for long-ranged systems with varying exponent $\alpha$. 
To demonstrate the universality of the features we observe, we also perform analogous quench procedures on fermions, by employing the fermionic variant of the MCTDH method (MCTDHF).
Both methods are implemented in the software MCTDH-X~\cite{Alon:2008,Lode:2016,Fasshauer:2016,Lin:2020,Lode:2020,MCTDHX,ultracold}.
MCTDH-X has been a very prolific tool for studying the dynamics of long-range interacting quantum gases in the past~\cite{Lode:2012,Cao:2013,barnali_axel, Fischer:2015,chatterjee:2018,Molignini:2018,Bera:2019,Lin:2019,chatterjee:2019,chatterjee:2020,Lin:2020-PRA,Lin:2021,Molignini:2022,Rosa-Medina:2022,Hughes:2023,Molignini:2024,Bilinskaya:2024,Molignini:2024-2}, as it enables us to evaluate the full time evolution of time-dependent processes in a very efficient way.

In our simulations, we find that the time evolution exhibits violent dynamical fragmentation, unbounded entropy production, and complete delocalization in configuration space.
These three features suggest that the highly excited post-quench phase behaves as a quasiclassical gas when starting from a crystal state exhibiting long-range interactions.
From the analogous simulations performed on fermions, we observe a very similar behavior which indicates the universality of our results in the dynamics of non-local interacting systems.
For both bosons and fermions we observe very exotic dynamics although the system eventually relaxes to a fully occupied configuration space for the entire range of $\alpha$.
Somewhat surprisingly, we do not find any signatures of any intermediate time scale in the relaxation process.
However, we observe an exponential boundary in the entropy evolution between relaxed and non-relaxed states, indicating a progressive slowdown in the relaxation process at increasing longer-ranged interactions (lower value of $\alpha$). 
To understand the process of relaxation, the system is also studied for quenches to other weakly attractive regimes. 
We observe that the relaxation process is less violent, slows down and begins to exhibit a quadratic entropy scaling with a very slow growth at short times when $\alpha<1$.

The present study sheds light on dynamical universal behavior of highly non-local interacting systems, which directly relates to the large amount of research in state-of-the-art experiments in the AMO systems. 
We envision a rich interplay between long-range interactions and post quench dynamics both for ultracold bosons and fermions. 
We believe that our in-depth analysis can be exploited to understand the richness of these versatile quantum long-range systems.

The paper is structured as follows. 
In Sec. II, we discuss the Hamiltonian and theoretical approach. 
In Sec. III, we introduce quantities of interest. Sec. IV discusses the coherent to chaotic behavior through the analysis of three key quantities: orbital occupation, Fock space occupation, and entropy measures. 
Sec. V illustrates the underlying relaxation process both for bosons and fermions. 
Sec. VI presents how to control the quench mechanism to achieve a drastic dynamics slowdown.
We conclude our paper in Sec VII. 
Appendix A presents an overview of the initial states, Appendix B considers a stronger quench dynamics, and Appendix C summarizes the units used in the numerics.

%%%%%%%%%%%%%%%%%%%%%%%%%%%%%%%%%%%%%%%%%%%%%%%

\section{Model and theoretical approach}
The time evolution of $N$ interacting bosons is governed by the TDSE (here and henceforth, we set $\hbar=1$)
\begin{equation}
i \frac{\partial \psi}{\partial t} = \hat{H} \psi,
\end{equation}
where the total Hamiltonian for the system is
\begin{equation} 
\hat{H}(x_1, \dots, x_N)= \sum_{i=1}^{N} \hat{h}(x_i) + \sum_{i<j=1}^{N}\hat{W}(x_i - x_j)
\label{propagation_eq}
\end{equation}
The Hamiltonian $\hat{H}$ is expressed in dimensionless units
Please refer to appendix C for a complete discussion of how to obtain dimensionless units in MCTDH-X~\footnote{The Hamiltonian $\hat{H}$ is obtained by dividing the dimensionful Hamiltonian by $\frac{\hbar^{2}}{mL^{2}}$, with $m$ the mass of the particles and $L$ an arbitrary length scale, which we choose to be the point at which the harmonic trap equals $\frac{1}{2}$.}.
The operator $\hat{h}(x) = \hat{T}(x) + \hat{V}(x)$ is the one-body Hamiltonian containing the kinetic energy $\hat{T}(x)=-\frac{1}{2} \hat{\partial}_x^{2}$ and the external potential $\hat{V}(x)=\frac{1}{2} x^2$.
The operator $\hat{W}(x_i - x_j)$ describes the two-body interaction between particles at positions $x_i$ and $x_j$ (which can be either attractive or repulsive).

The MCTDH ansatz for the bosonic many-body wave function is a linear combination of time dependent permanents constructed over $M$ single-particle wave functions called orbitals.
The ansatz for the fermionic many-body wave function is equivalent, but Slater determinants replace the permanents. 
We redirect to Ref.~\cite{Lode:2016} for further details.
Both cases can be written as
\begin{equation}
\vert \psi(t)\rangle = \sum_{\bar{n}}^{} C_{\vec{n}}(t)\vert \vec{n};t\rangle.
\label{eq:many_body_wf}
\end{equation}
The vector $\vec{n} = (n_1,n_2, \dots ,n_M)$ represents the occupation of each orbital with the constraint that $n_1 + n_2 + \dots +n_M = N$, which ensures the preservation of the total number of particles. 
Distributing $N$ bosons over $M$ time dependent orbitals, the number of configurations becomes 
\begin{equation}
N_{\mathrm{conf}} =  \left(\begin{array}{c} N+M-1 \\ N \end{array}\right).
\end{equation}
\label{eq:N-conf}
It is important to emphasize that in the ansatz~\eqref{eq:many_body_wf}, both the expansion coefficients  $C_{\vec{n}}(t)\vert \vec{n};t\rangle$ and the orbitals that build up the configurations $\vert \vec{n};t\rangle$ are time-dependent, fully variationally optimized quantities.
This time-adaptive basis allows the sampled Hilbert space to dynamically follow the motion under any quench dynamics. 

In the limit $M \rightarrow \infty$, the representation of Eq.~\ref{eq:many_body_wf} is exact.
However, in practice, the size of the Hilbert space has to be truncated by using a finite value of $M$ during computation.
The time-dependent orbitals assure that a given degree of accuracy is reached with a much shorter expansion compared to the time independent basis used in exact diagonalization. 

The occupation of the different orbitals offers a measure for the many-body state dynamical fragmentation in the nonequilibrium dynamics of the quantum quench.
More precisely, we define fragmentation from the natural occupations $n_i$, i.e. the population of the natural orbitals, which are the eigenvalues of the reduced one-body density matrix $\rho^{(1)}(x,x';t) = \langle \psi(t) \vert \hat{\psi}^{\dagger}(x') \hat{\psi}(x) \vert \psi(t) \rangle$, i.e.
\begin{equation}
\rho^{(1)}(x, x', t) = \sum_i n_i \phi_i^*(x') \phi_i^*(x) 
\label{eq:rho1}
\end{equation}
as a spectral decomposition.
In the expression for the reduced one-body density matrix, $\hat{\psi}^{\dagger}(x)$ and $\hat{\psi}(x)$ are field operators for the creation and annihilation of one particle at position $x$.
For a non-fragmented or condensed state, the occupation of the lowest natural orbital is close to unity. 
However, for a fragmented state, several natural orbitals may contribute significantly.  
In particular, for bosons, (lack of) fragmentation illustrates how close the many-body state is to a global superfluid that can be described by a single wave function.

\section{Quantities of interest}
\label{sec:quantities}
To study statistical relaxation, the Shannon information entropy (SIE) $S^{\mathrm{info}}(t)$ is the ideal quantity~\cite{Berman:2004}. 
The SIE of the one-body density in coordinate space is defined as 
$S^{\mathrm{info}}(t) = -\int { \mathrm{d}x \rho^{(1)}(x,t) \ln \left[{\rho^{(1)}(x,t)} \right] }$.
The SIE is a measure of localization/delocalization of the corresponding density. 
However, it is insensitive to the correlation present in $\vert \psi(t) \rangle$.
Considering the MCTDH ansatz~\eqref{eq:many_body_wf}, then, we define a few alternative entropy measures~\cite{barnali_axel}.

We begin by considering information that can be extracted from the coefficients of the MCTDH expansion.
The modulus squared of each coefficient can be expressed as 
\begin{equation}
|C_{\vec{n}}(t)|^2 = \frac{1}{\prod_{i=1}^M n_i!} \left< \Psi \right| (\hat{b}_1(t))^{n_1} \cdots (\hat{b}_M(t))^{n_m} (\hat{b}_1^{\dagger}(t))^{n_1} \cdots (\hat{b}_M^{\dagger}(t))^{n_m} \left|\Psi \right> 
\end{equation}
indicating that -- depending on the state -- one, some, or all $M$ creation and annihilation operators may contribute to the value of the coefficient $|C_{\vec{n}}(t)|^2$. 
Consequently, the distribution of these coefficients provides a direct qualitative assessment of the many-body entropies within the system.

The \emph{coefficient Shannon information entropy} is defined as 
\begin{equation}
 S_{C}(t) = -\sum_{\vec{n}} | C_{\vec{n}}(t)|^2 \ln |C_{\vec{n}}(t)|^{2}.
\end{equation}
It characterizes the distribution of the state $\vert \psi(t) \rangle$ in the Fock space.
A related quantity is the \emph{coefficient inverse participation ratio}, defined as
\begin{equation}
    I_C(t) = \frac{1} {\sum_{\vec{n}}| C_{\vec{n}}(t)|^4}.
\end{equation}
This is another measure to detect the effective number of basis states participating in the time evolution of the many-body state and is generally utilized to understand irregular dynamics. 
We can also define a hybrid version of the coefficient Shannon information entropy, the \emph{$N$-body coefficient entropy} $S_C^N(t)$, as
\begin{equation}
S_{C}^N(t) = -\sum_{\vec{n}, \vec{n^{\prime}}} | C_{\vec{n}}(t)|^2 \ln |C_{\vec{n^{\prime}}}(t)|^{2}.
\end{equation}
This quantity also conveys information about delocalization in Fock space, but considering all possible pairings of configurations.
In terms of properties, it inherits most of its features from the coefficient entropy $S_C(t)$. 
Therefore, in the following, we will focus our discussion on $S_C(t)$ and $I_C(t)$.

We highlight a few limiting cases for these two quantities.
When the system is (and remains) in a single-configuration state, $S_C(t)=0$ and $I_C(t)=1$.
This type of state is equivalent to a mean-field configuration.
By definition, then, mean-field approaches such as the Gross-Pitaevskii equation or multiorbital methods will \emph{always} lead to a trivial coefficient entropy and inverse participation ratio.
In MCTDHB, instead, these quantities can acquire much larger values.

At the other extreme, when the $N$-body Fock space is completely and uniformly populated, $S_C(t)$ and $I_C(t)$ will saturate to their maximal values.
This situation indicates that the MCTDHB expansion has exhausted all the available configuration space to describe the many-body state. 
This is typical of systems with strong interactions or significant entanglement, where the particles do not behave independently but are instead highly correlated.
Another example is chaotic systems which sample the entire configuration space. 
Strictly speaking, this situation can occur only in the limit $M \to \infty$.
If $S_C(t)$ and $I_C(t)$ show an unbound increase when evaluated for progressively larger configuration space (larger values of $M$), it is an indication of chaotic-like or quasi-classical (incoherent) behavior.

At intermediate values of $S_C(t)$ and $I_C(t)$, $\vert \psi(t) \rangle$ spreads over only a part of the available configuration space and several but not all expansion coefficients are nonzero. 
In this case, the coefficient entropy gives a quantitative measure of the many-body character of the state at time $t$, which interpolates between the fully coherent, single-configuration case, and the fully chaotic, infinite-configuration case.
Similarly, when $I_C(t)$ is small, the corresponding many-body state is close to the mean-field state and the corresponding dynamics is regular. 
Large $I_C(t)$ indicates instead that the corresponding dynamics is irregular or chaotic.

In a quench procedure like the one we study in this work, we typically start with a moderately correlated state exhibiting a low or intermediate value of $S_C(t)$ and $I_C(t)$, and we then observe their time evolution.
If a lot of energy is injected into the system, $S_C(t)$ and $I_C(t)$ will typically increase in time, indicating spread of correlations, Fock-space delocalization, and progress towards a chaotic state.

While the entropies introduced above give a good grasp of the many-body character of the state and its potential progress towards a chaotic configuration, they are based on coefficients alone, which are not uniquely defined and instead depend on the many-body state decomposition into orbitals (i.e. the permanents). 
Rotating the orbitals by a unitary transformation can be compensated by transforming the coefficients. 
Thus, any MCTDH wavefunction has infinite representations: $\left| \Psi \right> = \sum_{\vec{n}} C_{\vec{n}} \left| n \right> =  \sum_{\vec{m}} C_{\vec{m}} \left| m \right>$ etc, i.e. the same many-particle wavefunction can thus be described by different orbitals and different coefficients. 
While the MCTDH variational procedure converges to a state with the most optimized natural occupation, and this correlates with a restricted configuration population, there is no guarantee that the achieved configuration will be ``minimal'' in some sense. 
There is a way of selecting coefficients with the narrowest distribution based on variance calculations~\cite{Streltsov:2006}, but they are quite cumbersome and deserve a standalone treatment, which is beyond the scope of our study.

To obviate to the lack of invariance for the coefficients, we thus consider another measure of entropy obtained from explicitly invariant quantities, namely the natural occupations defined in Eq.~\ref{eq:rho1}.
This quantity is called \emph{occupation Shannon information entropy} and is defined as 
\begin{equation}
S_n(t)= - \sum_{i} \bar{n}_i(t) [\ln \bar{n}_i(t)],
\end{equation}
where $\bar{n}_i(t) = \frac{n_i(t)}{N}$ are natural occupations normalized with the particle number.

This entropy quantifies the distribution of particles among the various natural orbitals. 
A lower occupation entropy indicates that particles are predominantly occupying fewer orbitals (signifying higher coherence), whereas a higher entropy suggests a more uniform distribution across multiple orbitals (indicative of fragmentation and reduced coherence).

The limiting cases for this quantity are similar to those of the coefficient entropy.
In a single-orbital mean-field theory, as the reduced density matrix has only a single eigenvalue, $S_n(t)=0$ and the corresponding many-body state is non-fragmented. 
At the opposite extreme, when $M \to \infty$ and all the orbitals have a finite (uniform) occupation, we are in a completely chaotic regime.
In the intermediate cases, when multiple significant orbitals contribute but the occupation is not uniform at $1/M$, the corresponding many-body state is fragmented but retains a certain degree of coherence.

In a relaxation process, $S_n(t)$ initially starts with a minimum value (which may not be zero), increases, and finally saturates to a maximal value. 
Therefore, $S_n(t)$ conveys the emergence and degree of fragmentation over time.
If $S_n(t)$ appears to increase unboundedly when the same simulation (same initial conditions) is performed but with an increasing number of orbitals, it is a sign that the system relaxes towards a chaotic state where all occupations are macroscopic, i.e. all the eigenvalues of the reduced one-body density matrix have the same magnitude.

We remark that even for multiorbital mean-field theories, which can exhibit a finite value of $S_n(t)$, this value stays constant throughout the dynamics because the populations cannot change.
Thus, this quantity is a prerogative of multiconfigurational methods like MCTDHB only.

\section{Quantum to classical behavior}
We now present the results of our quench dynamics calculations.
We consider generalized long-range (power law) interactions determined by the exponent $\alpha$:
\begin{equation}
    \hat{W} (x_i - x_j) =\frac{g_d}{\vert x_i - x_j \vert ^{\alpha} + \epsilon_{\alpha}}.
\end{equation}
The parameter $g_d$ controls the strength of the long-range interactions. 
In the current platform of generalized AMO systems, $\alpha$ is scaled according to the type of interaction and generally the ratio $\frac{\alpha}{d}$ measures how strong the interaction is. 
For our present study in 1D system, $\alpha$ simply measures the range of interaction considering both weak long-range and strong long-range interactions. 
In general, $\alpha=1$ appears for gravitational or Coulomb interactions, $\alpha = 3$ describes a dipolar quantum gas, and $\alpha=0$ corresponds to interaction mediated by cavity photons (infinite ranged). 
For $\alpha=3$ (dipole-dipole interactions), $g_d$ could be expressed as $g_d = \frac{d_m^{2}}{4\pi\epsilon_0}$ for electric dipoles and $g_d = \frac{d_m^2\mu_0} {4\pi}$ for magnetic dipoles, $d_m$ being the dipole moment, $\epsilon_0$ is the vacuum permittivity, and $\mu_0$ is the vacuum permeability.
In general, the dipole-dipole interaction potential in 1D also includes a contact term owing to the transverse confinement, however that can be safely neglected for strong interaction strengths~\cite{Budhaditya}.
Unless otherwise stated, in the quench procedure we employ $g_d = 100.0 \to -100.0$.
The parameter $\epsilon_{\alpha}$ is a short-range cutoff to avoid unphysical singularities arising at $x_i = x_j$. 
Note that $\epsilon_{\alpha}$ is varied according to the choice of $\alpha$. 
We tune $\alpha$ in the range $\left[ 0.5, 4 \right]$, which considers many different kinds of generalized long-range quantum systems. 
We choose the cut-off parameter such that the effective interaction $V_{\mathrm{eff}}$= $\int_{\mathcal{D}} {\frac{g_d} {x^{\alpha} + \epsilon_{\alpha}} \mathrm{d} x} = \int_{\mathcal{D}} { \delta(x) \mathrm{d} x} = 1 $, where the $\mathcal{D} = [-15,15]$ encompasses the entire simulation domain.

Before assessing the process of statistical relaxation for general $\alpha$, it is instructive to discuss the Fock space dynamics of the excited metastable attractive phase for a particular case. 
We select $\alpha$ = 3.0, and corresponding $\epsilon_{\alpha=3.0}$  $\approx 0.1015976$. 
To understand the dynamical fragmentation during the quench, we perform a study with increasing orbital number $M$ and present results for the selected values of $M= 5, 10, 15, 20$. 

\subsection{Violent fragmentation}
\begin{figure}
    \centering
    \includegraphics[width=\columnwidth, angle=0]{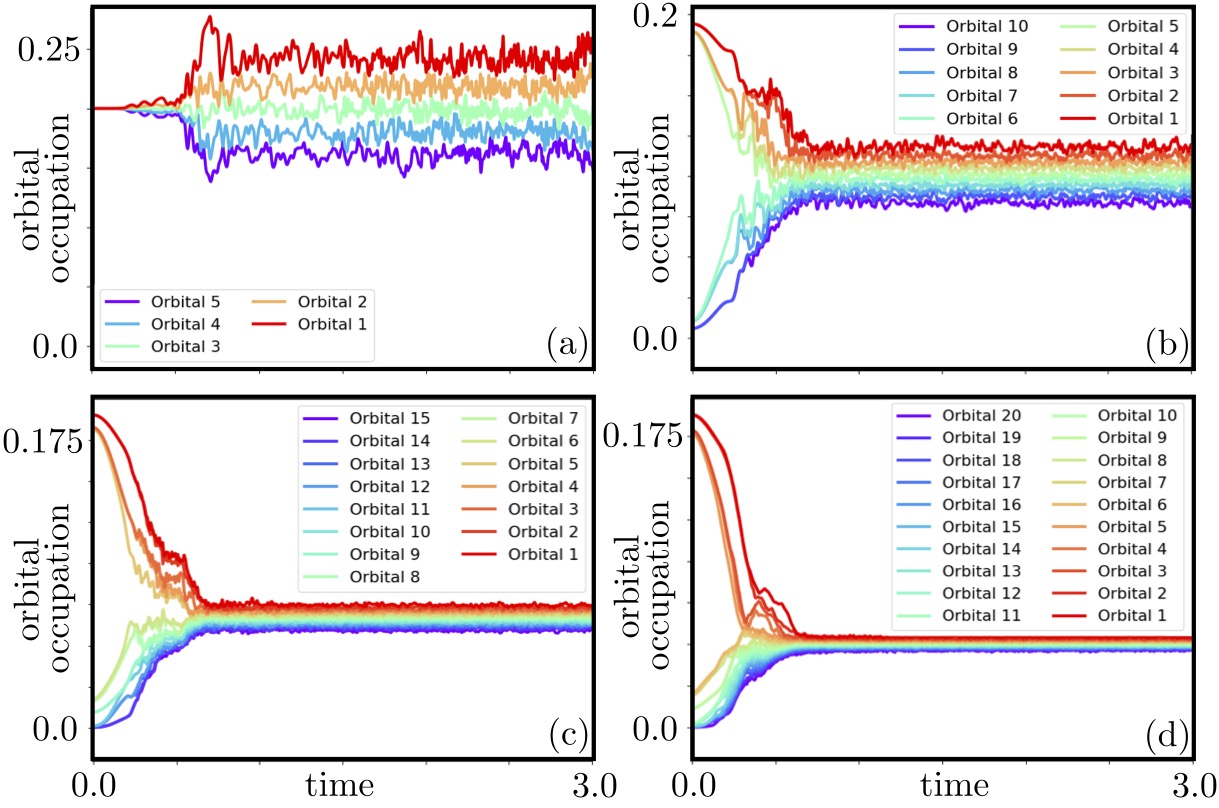}
    \caption{Orbital fragmentation dynamics for $N=5$ quenched bosons with $\alpha = 3.0$, $g_d=100.0 \to -100.0$ described with (a) $M=5$, (b) $M=10$, (c) $M=15$, (d) $M=20$ orbitals.}
    \label{fig:fragmentation}
\end{figure}

Fig.~\ref{fig:fragmentation} depicts the time evolution of the orbital occupation.
The figure shows that initially $(t < 0)$ the system is fragmented for all choices of orbital number. 
The occupation in the natural orbitals changes quite rapidly after the quenching procedure. 
The occupation in the orbitals with initial significant population decreases, while the one in the initially nearly empty orbitals increases.
This trend continues until a saturation point, located approximately at time $t=0.5$ for all values of $M$, where the population of the orbitals coalesces around the value $1/M$.
This represents an irregular behavior which we term {\it violent fragmentation}, and is the hallmark of a complete saturation of the probed Hilbert space of our quantum simulations.
We remark that this type of violent fragmentation is exhibited --- practically unchanged --- for all values of $M$ up to $M=20$, which represents a computational upper threshold for our simulations given the current state-of-the-art.
At the transition time the fully collapsed state always occupies the full available configuration space.
The quench from strongly repulsive to strongly attractive long-range interactions can therefore be described as a transition process from {\it quantum to classical} behavior of the long-range interacting bosons. 

%%%%%%%%%%%%%%%%%%%%%%%%%%%%%%%%%%%%%%%%
%%%%%%%%%%%%%%%%%%%%%%%%%%%%%%%%%%%%%%%%

\subsection{Fock space delocalization}
\begin{figure}
    \centering
    \includegraphics[width=\columnwidth, angle=0]{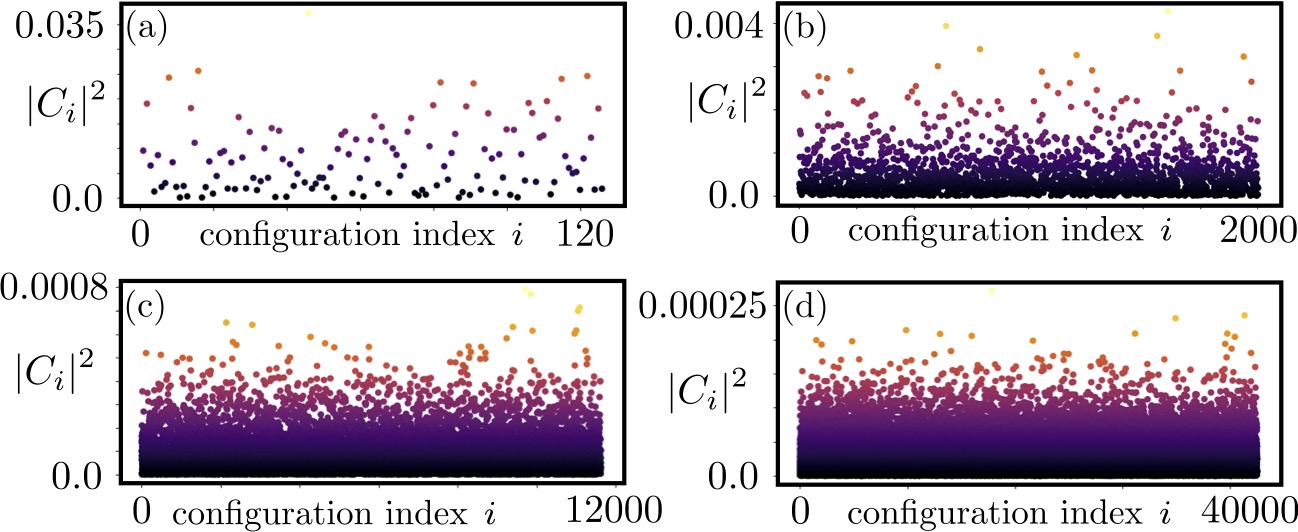}
    \caption{Configuration space population at time $t=3.0$ as measured from the amplitude squared of the many-body coefficients, for $N=5$ quenched bosons with $\alpha = 3.0$, $g_d =100.0 \to -100.0$ and increasing orbital number $M$. The orbital number is (a) $M=5$, (b) $M=10$, (c) $M=15$, (d) $M=20$.}
    \label{fig:coefficients}
\end{figure}

To further understand the quantum to classical behavior, we visualize the value of the coefficients $C_{\bar{n}}(t)$ after the quench in Fig.~\ref{fig:coefficients}
We plot $|C_{\bar{n}}(t)|^2$ as a function of the index $n$ of the basis states for the final simulation time at $t=3.0$.
The number of configurations in the available space for particular choice of $M$ is determined by $N_{\mathrm{conf}}$ by Eq.(4).
Thus, $N_{\mathrm{conf}}$ increases exponentially from 126 for $M=5$ to over 40,000 for $M=20$.
For each choice of $M$, the initial TG gas undergoes a violent delocalization process in which most of the configurations acquire a comparable weight.
This behavior corresponds to the definition of classical gas, which occupies the whole available configuration space. 

%%%%%%%%%%%%%%%%%%%%%%%%%%%%%%%%%%%%%%%%
%%%%%%%%%%%%%%%%%%%%%%%%%%%%%%%%%%%%%%%%

\subsection{Unbounded Entropy measures}
In the dynamical evolution of isolated quantum system, statistical relaxation is established by some kind of equilibrium. 
For interacting many-body systems, statistical relaxation is also related to chaos in the energy spectra~\cite{Santos:2012}.
In particular, chaos in quantum interacting systems is defined by the statistics of the eigenstates. 
The onset of chaos is established by quadratic growth in entropy at very short time, then linear growth before saturation to a maximum entropy states. 
For time reversal and rotationally invariant systems, the maximum entropy is also determined by the predictions of a Gaussian orthogonal ensemble of random matrices(GOE)~\cite{Kota, Rigol,Izrailev:2012,Mark:1994}. 

\begin{figure}
    \centering
    \includegraphics[width=\columnwidth, angle=0]{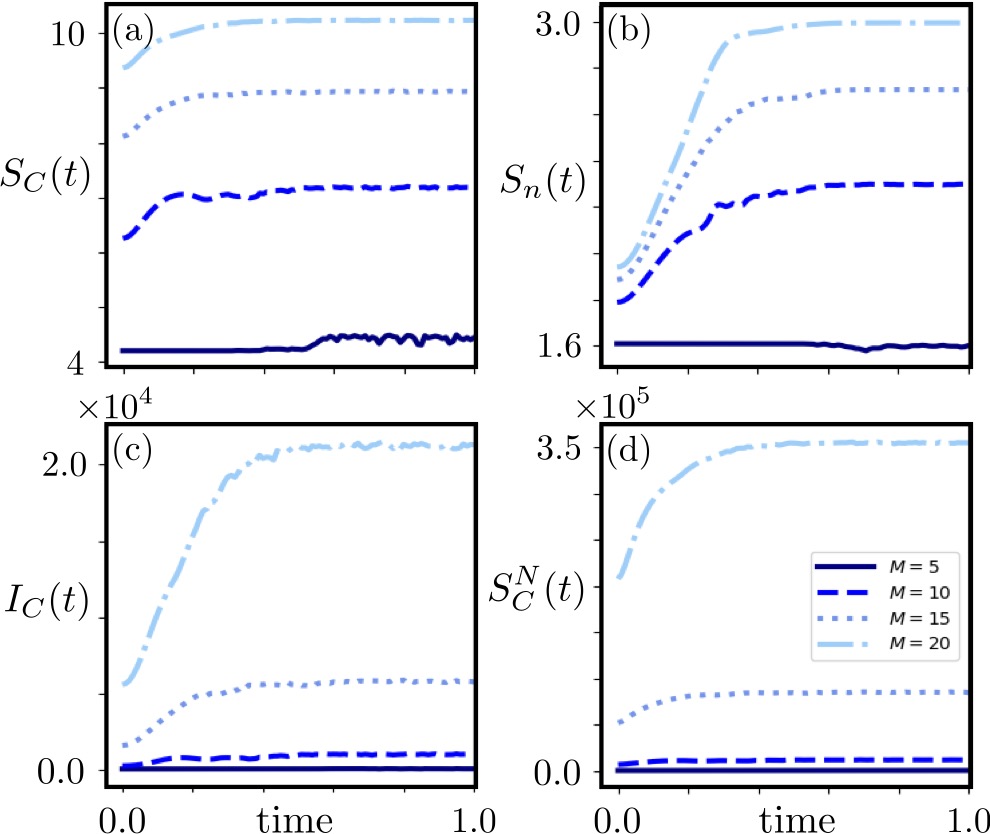}
    \caption{Dynamics for four different entropy measures for $N=5$ quenched bosons with $\alpha = 3.0$, $g_d=100.0 \to -100.0$ and increasing orbital number $M=5$ to $M=20$. (a) Coefficient entropy $S_C(t)$, (b) occupation entropy $S_n(t)$, (c) coefficient inverse participation ratio $I_C(t)$, (d) many-body coefficient entropy $S_C^N(t)$.}
    \label{fig:unbounded-entropy}
\end{figure}

Therefore, to further understand the violent dynamics and its impact on the relaxation process, in Fig~\ref{fig:unbounded-entropy} we plot the four different entropic quantities introduced in section \ref{sec:quantities} ($S_C(t)$, $S_n(t)$, $I_C(t)$ and $S_C^N(t)$) as measures of irregular dynamics.
We present how the different measures approach the GOE prediction as determined by the number of orbitals used in the computation. 
For a GOE of random matrices, $I_C^{GOE}=D/3$ and $S_C^{GOE}=\ln 0.48 D$, where $D\times D$ is the dimension of the random matrices~\cite{Kota}.
To compare with GOE estimates, we set $D$ as equal to the number of configurations $N_{\mathrm{conf}}$ participating in the dynamics. For occupation entropy $S_n$, $D$ is set equal to the number of orbitals $M$ used in the computations. 
Thus $S_n^{GOE}$ = $-\sum_{i=1}^{M} \frac{1}{M} \ln(\frac{1}{M}) = \ln(M)$. 
The GOE estimates are evaluated for different $M$ used in the MCTDH-B treatment.  

Fig.~\ref{fig:unbounded-entropy} shows that statistical relaxation is achieved simultaneously with Fock space delocalization and violent dynamical fragmentation at time $t$=$0.5$.
In particular, the results for increasing orbital number further support the thesis that the production of entropy is indeed {\it unbounded}.
For example, for $M=15$, the number of configurations in the dynamics is $N_{conf}$ = 11'628.
The corresponding GOE predictions are  $S_{C}^{GOE}$ = 8.627, $S_n^{GOE}=2.708$, $I_C^{GOE}$= 3'876. 
We observe that for the prequench state $(t=0)$, all the measures are far enough from $0$ indicating that the system initially is fragmented and delocalized. 
With time, all the entropy measures smoothly increase and then saturate. 
The saturation values for $M=15$ are $S_C^{\mathrm{sat}} \approx 8.94$, $S_n^{\mathrm{sat}} \approx 2.71$ and $I_C^{\mathrm{sat}}= 5'820$. 
These values are indeed close to GOE predictions, however a slight discrepancy exists because the interaction strength needed to exhibit a TG-like limit in the initial state is very large, albeit not infinite, which is the perquisite for GOE prediction. 
Repeating the same calculations with $M=20$ yields $N_{\mathrm{conf}}$=42'504, and the corresponding GOE predictions are $S_{C}^{GOE}$ = 9.923, $S_n^{GOE}=2.995$, $I_C^{GOE}$= 14'168. 
The corresponding saturation values in MCTDH-B computations are  $S_C^{\mathrm{sat}} \approx 10.24$, $S_n^{\mathrm{sat}} \approx 2.99$ and $I_C^{\mathrm{sat}} \approx 21'372$. 
The saturation close to GOE prediction  for progressively larger values of $M$ (and hence the configuration space) indeed exhibits the hallmark of statistical relaxation. 
However, the impact of violent fragmentation is visible in the increase of the entropy measures with the number of orbitals.
These combined results suggest that the entropy can increase without bound with increase of the orbital number, similarly to how a classical gas occupies the entire available configuration space when unconstrained.

%%%%%%%%%%%%%%%%%%%%%%%%%%%%%%%%%%%%%%%%
%%%%%%%%%%%%%%%%%%%%%%%%%%%%%%%%%%%%%%%%

\section{Understanding of relaxation process}
Understanding the relaxation process is an important unsolved problem in interacting quantum many-body systems. 
Theoretical concepts like the quantum ergodic theory or the eigenstate thermalization hypothesis (ETH)~\cite{Mark:1994} infer  possible relaxation processes described through generalized Gibbs ensembles. 
However, it is unclear on what time scale this occurs depending on the interaction range.
It has been conjectured that several distinct time scales can arise in the dynamics of complex systems, although eventually they do settle to an equilibrium state in the long time dynamics~\cite{Langen:2015}.

An intriguing question to understand is the intermediate path from a relaxing state to a fully relaxed one. 
In this section, we thus analyze the many-body quantum dynamics over the entire range of the exponents $\alpha \in \left[0.5,4\right]$. 
To understand the relaxation process we thoroughly discuss two typical cases: $\alpha=2.7$ (weakly long-range) and $\alpha=0.7$ (strongly long-range). 
Besides thoroughly investigating the dynamics of bosons, we also present calculations for spinless fermions to establish the universality of the fragmentation and relaxation processes. 
For both quantum statistics, we maintain the same protocol in the quench dynamics. 
We finally summarize the entire results in a parameter-time diagram which depicts the phase boundary between non-relaxed gas and relaxed gas.

\subsection{Process of relaxation for bosons}
We begin by probing the relaxation process for bosons.
As before, we consider a sudden quench from $g_d=100.0 \to -100.0$ with a fixed number of particles $N=5$, and concentrate on the results for $M=10$ orbitals.

\begin{figure}
    \centering
    \includegraphics[width=\columnwidth, angle=0]{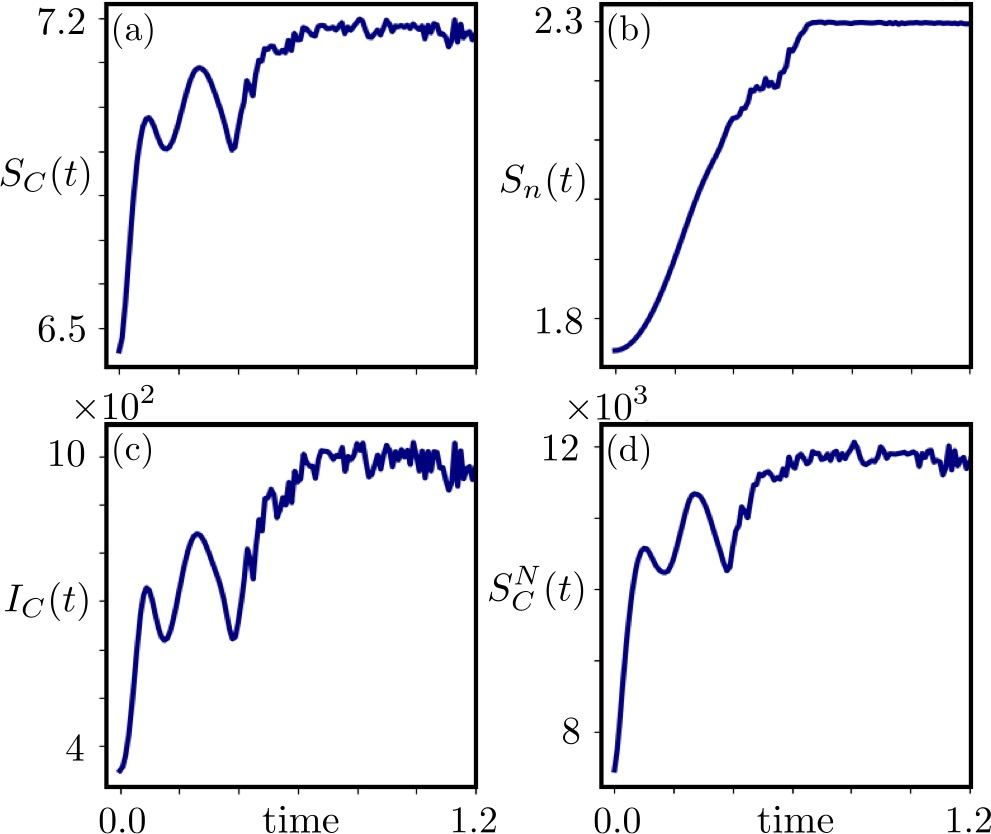}
    \caption{Dynamics for four different entropy measures for $N=5$ quenched, very long-range interacting bosons with $\alpha = 0.7$, $g_d=100.0 \to -100.0$, $M=10$. (a) Coefficient entropy $S_C(t)$, (b) occupation entropy $S_n(t)$, (c) coefficient inverse participation ratio $I_C(t)$, (d) many-body coefficient entropy $S_C^N(t)$.}
    \label{fig:entropy-bosons-alpha-0.7}
\end{figure}

First, we discuss the case for strong long-range interactions with $\alpha=0.7$ and measure all the four entropic quantities $S_C(t)$, $S_n(t)$, $I_C(t)$ and $S_C^N(t)$, which are plotted in Fig.~\ref{fig:entropy-bosons-alpha-0.7}.  
We observe a concurring estimate from all the four different measures that the relaxation happens around time $t \approx 0.6$.
As expected, the saturation values approach the corresponding GOE predictions. 
Both $S_c(t)$ and $S_C^N(t)$ exhibit peculiar oscillations before reaching the saturation value. 
This behavior suggests a highly complex dynamics with several intermediate and rather unstable phases before the entropy eventually settles to the final state. 
$S_n$, instead, exhibits an almost linear increase at short times and smoothly settles to the GOE value. 
$S_n$ is more insensitive to irregularities in the dynamics as it is determined by the fragmentation of the many-body states, which is a many-body invariant.
In fact, in the previous section we already saw that $S_n$ is the entropic quantity that best converges to the GOE result.

%%%%%%%%%%%%%%%%%%%%%%%%%%%%%%%%%%%%%%%%
%%%%%%%%%%%%%%%%%%%%%%%%%%%%%%%%%%%%%%%%

\begin{figure}
    \centering
    \includegraphics[width=\columnwidth, angle=0]{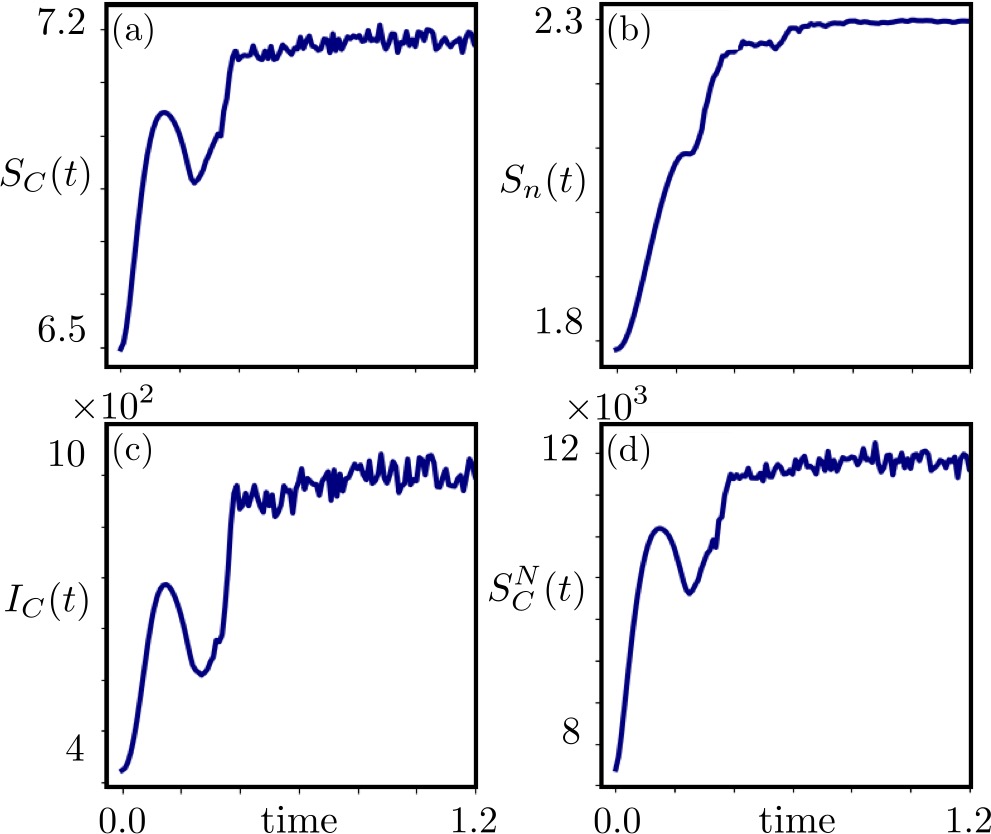}
    \caption{Dynamics for four different entropy measures for $N=5$ quenched, nearly dipolar-interacting bosons with $\alpha = 2.7$, $g_d=100.0 \to -100.0$, $M=10$. (a) Coefficient entropy $S_C(t)$, (b) occupation entropy $S_n(t)$, (c) coefficient inverse participation ratio $I_C(t)$, (d) many-body coefficient entropy $S_C^N(t)$.}
    \label{fig:entropy-bosons-alpha-2.7}
\end{figure}

The results for weak long-range interactions with $\alpha=2.7$ are presented in Fig.~\ref{fig:entropy-bosons-alpha-2.7}.  
The corresponding relaxation process is less complex than for strong long-range interactions.
In the intermediate dynamics, the entropy measures exhibit a single but deeper minimum which is superseded by the eventual saturation to the GOE value.
There is no clear signature of quasi equilibrium states for this case, and the time to reach the relaxed state is shorter ($t \approx 0.4$) than for strong long-range interactions. 

%%%%%%%%%%%%%%%%%%%%%%%%%%%%%%%%%%%%%%%%
%%%%%%%%%%%%%%%%%%%%%%%%%%%%%%%%%%%%%%%%

\begin{figure}
    \centering
    \includegraphics[width=0.7\columnwidth, angle=0]{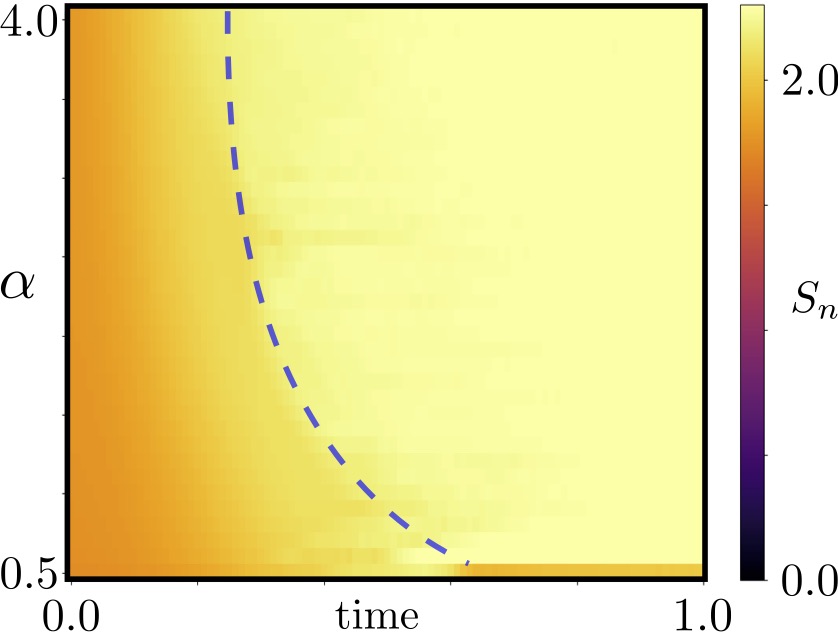}
    \caption{Occupation entropy dynamics as a function of interaction power $\alpha$ for $N=5$ quenched bosons with $g_d=100.0 \to -100.0$ and $M=10$.
    The dashed line indicates the exponential boundary between evolving and fully relaxed states.}
    \label{fig:PD-bosons}
\end{figure}

In Fig.~\ref{fig:PD-bosons}, we summarize our results in the form of a parameter-time diagram across the interaction power $\alpha$ and the simulation time.
Due to the closest agreement with GOE values for $S_n$ and its more regular dynamics that allows to determine the relaxation time more precisely, we will consider it as a good metric and use it as an order parameter in the parameter-time diagram.
The diagram shows a clear separation between non-relaxed and relaxed states and we observe an exponential boundary between these two regimes.
This suggests very long relaxation times in the limit $\alpha \to 0$, which cannot be reached with our current numerics due to the need of employing increasingly larger system sizes~\footnote{We remark that the reduction of entropy exhibited by $\alpha=0.5$ at $t>0.6$ is likely due to self-interference of the long-range tails of the interacting particles. In our numerics, although we employ a harmonic confinement, the system is always defined with periodic boundary conditions. If the system size is not large enough, as it might have happened for the $\alpha=0.5$ results, self-interference patterns across the system boundaries will appear and modify the dynamics.}.
The parameter-time diagram has more distinct characteristics when $\alpha <1$, i.e. when the range is less than the dimension of the system. 
For larger values, instead, the exponential boundary gradually loses its structure and the relaxation time is almost independent of the choice of $\alpha$.

%%%%%%%%%%%%%%%%%%%%%%%%%%%%%%%%%%%%%%%%
%%%%%%%%%%%%%%%%%%%%%%%%%%%%%%%%%%%%%%%%

\subsection{Process of relaxation for fermions}
\begin{figure}
    \centering
    \includegraphics[width=\columnwidth, angle=0]{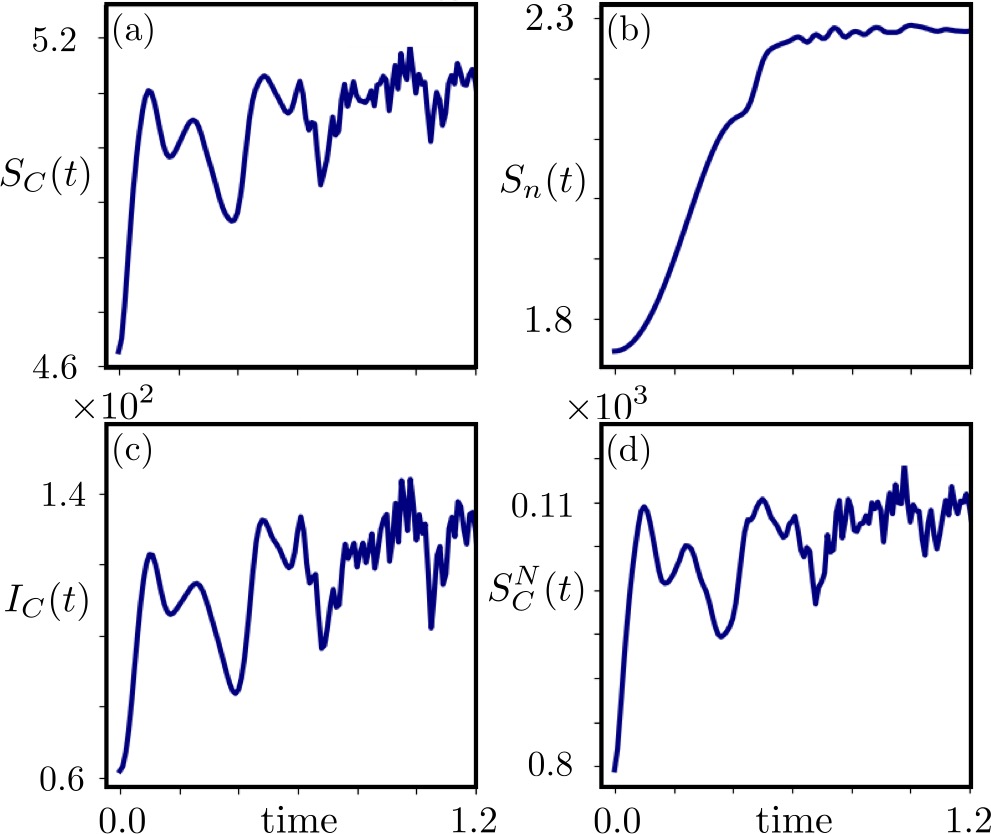}
    \caption{Dynamics for four different entropy measures for $N=5$ quenched, nearly dipolar-interacting fermions with $\alpha = 0.7$, $g_d=100.0 \to -100.0$, $M=10$. (a) Coefficient entropy $S_C(t)$, (b) occupation entropy $S_n(t)$, (c) coefficient inverse participation ratio $I_C(t)$, (d) many-body coefficient entropy $S_C^N(t)$.}
    \label{fig:entropy-fermions-alpha-0.7}
\end{figure}

\begin{figure}
    \centering
    \includegraphics[width=\columnwidth, angle=0]{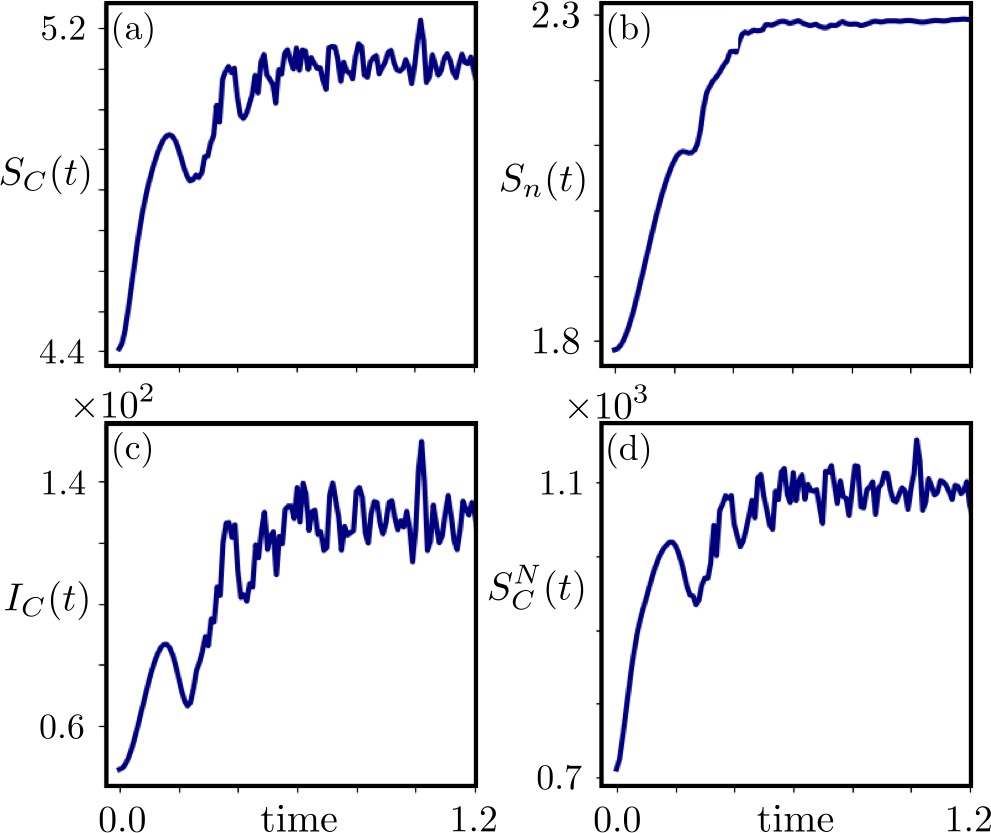}
    \caption{Dynamics for four different entropy measures for $N=5$ quenched, nearly dipolar-interacting fermions with $\alpha = 2.7$, $g_d=100.0 \to -100.0$, $M=10$. (a) Coefficient entropy $S_C(t)$, (b) occupation entropy $S_n(t)$, (c) coefficient inverse participation ratio $I_C(t)$, (d) many-body coefficient entropy $S_C^N(t)$.}
    \label{fig:entropy-fermions-alpha-2.7}
\end{figure}

In this section, we present similar many-body dynamics results for $N=5$ spinless \emph{fermions}, maintaining the same prequench and postquench parameters as chosen for the bosonic system. 
The corresponding measures of entropy are plotted in Fig.~\ref{fig:entropy-fermions-alpha-0.7} and Fig.~\ref{fig:entropy-fermions-alpha-2.7} for $\alpha = 0.7$ and $2.7$ respectively.  
In the strong long-ranged quench ($\alpha=0.7$), we observe highly modulated oscillatory structures in the dynamics of $S_C(t)$, $I_C(t)$ and $S_C^N(t)$. 
This makes it hard to determine the exact point of relaxation.
However, we find that the occupation entropy exhibits a smoother behavior and saturates to the same GOE prediction observed for bosons, which establishes the universality of the many-body dynamics in the quantum quench to the strong interaction regime.

%%%%%%%%%%%%%%%%%%%%%%%%%%%%%%%%%%%%%%%%
%%%%%%%%%%%%%%%%%%%%%%%%%%%%%%%%%%%%%%%%

For weak long-range interactions ($\alpha=2.7)$, oscillations with smaller amplitudes are observed, and the relaxation process becomes less complex. 
Eventual relaxation is observed in all the quantities and again $S_n$ agrees with the GOE prediction.

%%%%%%%%%%%%%%%%%%%%%%%%%%%%%%%%%%%%%%%%
%%%%%%%%%%%%%%%%%%%%%%%%%%%%%%%%%%%%%%%%

\begin{figure}
    \centering
    \includegraphics[width=0.7\columnwidth, angle=0]{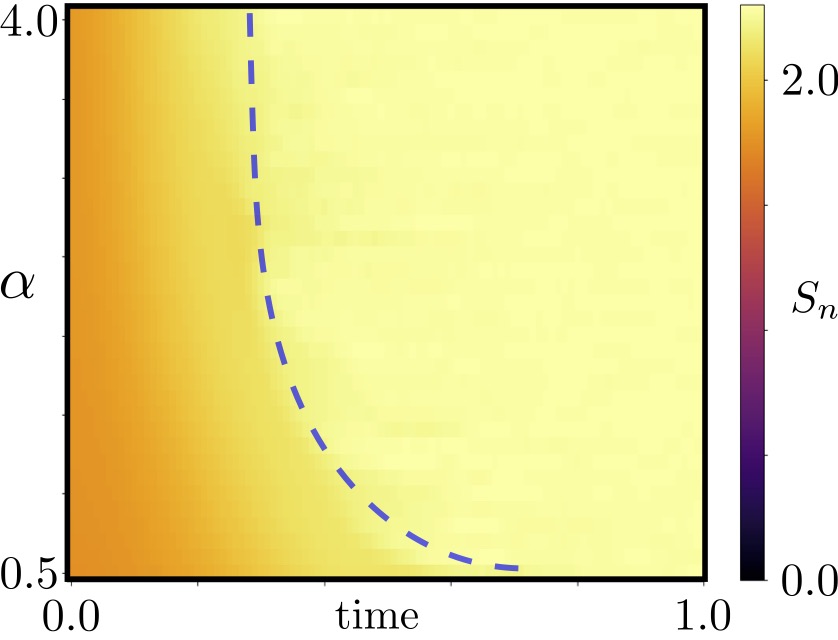}
    \caption{Occupation entropy dynamics as a function of interaction power $\alpha$ for $N=5$ quenched fermions with $g_d=100.0 \to -100.0$ and $M=10$.
    The dashed line indicates the exponential boundary between evolving and fully relaxed states.}
    \label{fig:PD-fermions}
\end{figure}

The fermionic phase diagram as shown in Fig.~\ref{fig:PD-fermions} exhibits an exponential boundary between the relaxing and relaxed states already observed for bosons.
However, the boundary is slightly shifted to larger times compared to the bosonic case when $\alpha <1$, whereas it remains rather unchanged for shorter range interactions.

%%%%%%%%%%%%%%%%%%%%%%%%%%%%%%%%%%%%%%%%
%%%%%%%%%%%%%%%%%%%%%%%%%%%%%%%%%%%%%%%%

\section{Control of relaxation} 
It is an established fact that a (near-)integrable system relaxes in at least two time scales related to the faster and slower relaxation processes~\cite{Langen:2015}.
Prethermalization occurs in the intermediate time scale before asymptotic relaxation to an equilibrium state. 
A necessary condition for thermalization is statistical relaxation in various observables to some kind of equilibrium.
However, the observed dynamics in the sudden quench probed here is a highly nonintegrable phenomenon. 
The dynamics is so complex that different intermediate states may exist, but we can not detect them due to their extremely small lifetimes. 
It is thus impossible to establish the possibility of any prethermal state in the strongly quenched regime within the current computational limitations.

\begin{figure}
    \centering
    \includegraphics[width=\columnwidth, angle=0]{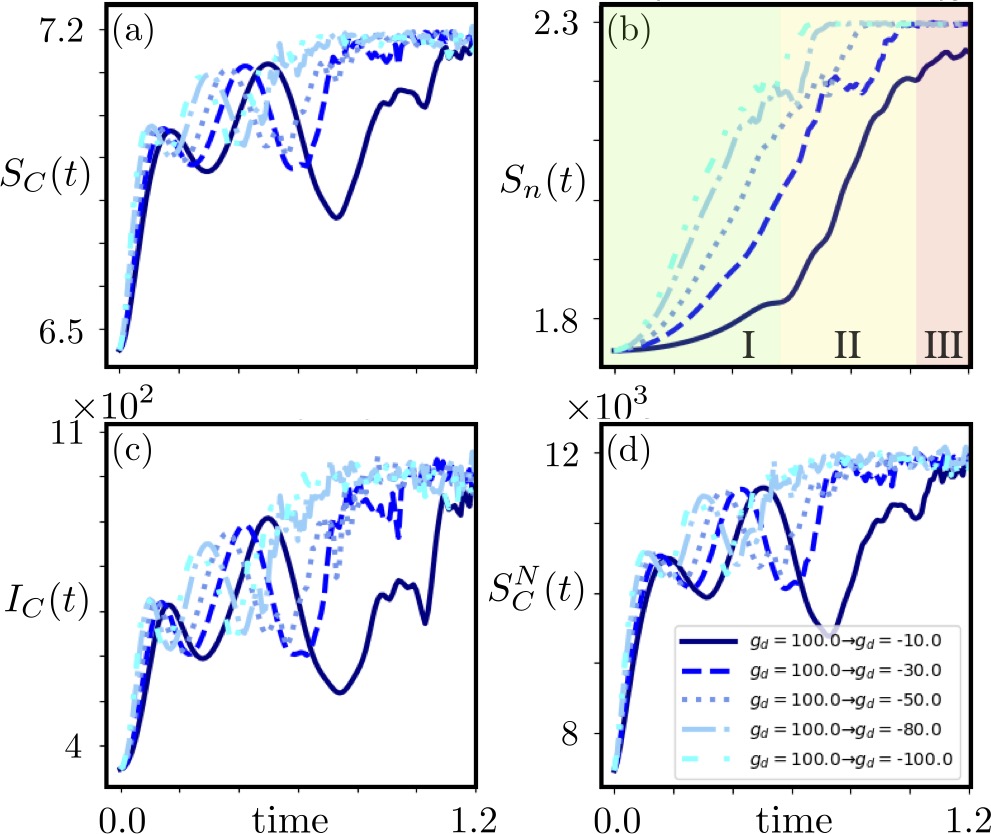}
    \caption{Dynamics for four different entropy measures for $N=5$ quenched, strongly long-range interacting bosons with $\alpha = 0.7$ and varying quench strength. (a) Coefficient entropy $S_C(t)$, (b) occupation entropy $S_n(t)$, (c) coefficient inverse participation ratio $I_C(t)$, (d) many-body coefficient entropy $S_C^N(t)$.
    For all curves $M=10$ orbitals were used.
    In panel (b), the three different background colors denote the three different regimes observed in the $g_d = -100.0 \to +10$ quench.}
    \label{fig:control-frag-alpha-0.7}
\end{figure}

\begin{figure}
    \centering
    \includegraphics[width=\columnwidth, angle=0]{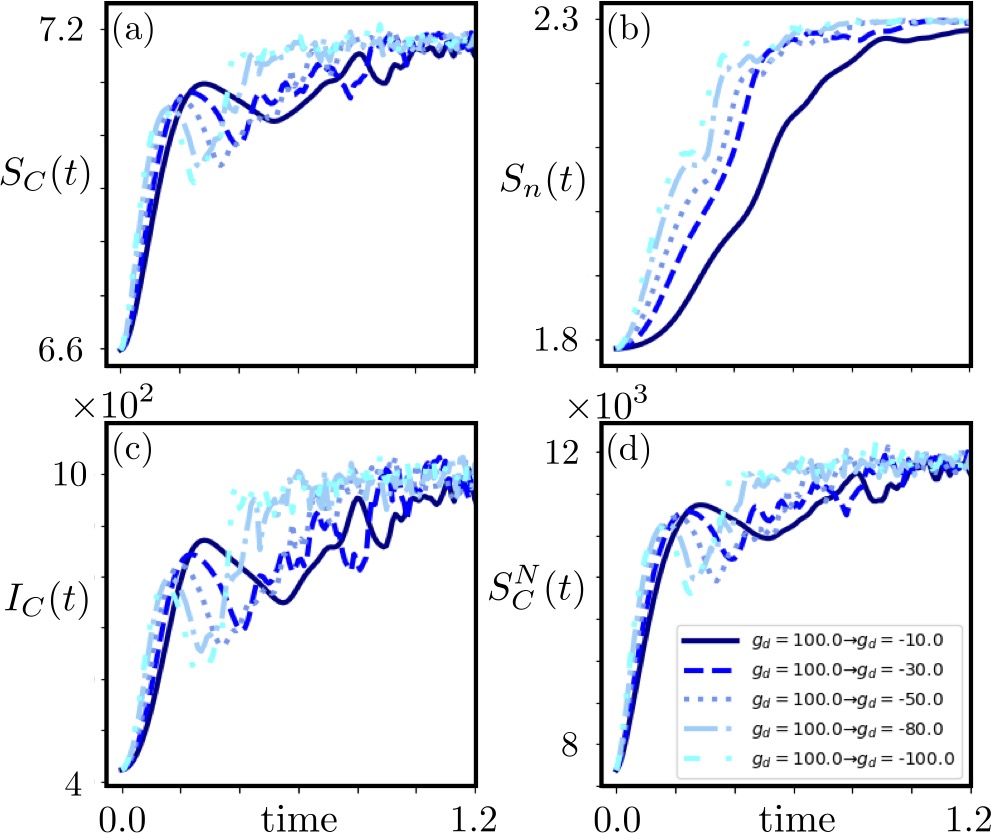}
    \caption{Dynamics for four different entropy measures for $N=5$ quenched, nearly dipolar-interacting bosons with $\alpha = 2.7$ and varying quench strength. (a) Coefficient entropy $S_C(t)$, (b) occupation entropy $S_n(t)$, (c) coefficient inverse participation ratio $I_C(t)$, (d) many-body coefficient entropy $S_C^N(t)$.
    For all curves $M=10$ orbitals were used.}
    \label{fig:control-frag-alpha-2.7}
\end{figure}

Nevertheless, another key question that can be addressed is whether it is possible to control the complex and extremely nontrivial pathways observed in the relaxation dynamics during the quench by probing regimes of weaker interactions.
In the following, we thus present an extended analysis of the dynamics for quenches with $g_d=+100.0$ to $g_d=-80.0, -50.0, -30.0$ and $-10.0$.
Again, we evaluate the measures of entropy dynamics for $N=5$ bosons for two distinct choices of $\alpha$. 
We present the results for $\alpha=0.7$ in Fig.~\ref{fig:control-frag-alpha-0.7} and for $\alpha=2.7$ in Fig.~\ref{fig:control-frag-alpha-2.7}. 
For $\alpha=0.7$, we observe that the dynamics is strongly affected by decreasing the interactions to weaker values.
The system exhibits more complex pathways and the relaxation process is severely hampered. 
In the same time scale, when $g_d$ is quenched from $+100.0$ to $-100.0$ and the system exhibits perfect relaxation after passing through many complex intermediate states, the $g_d$= $+100.0$ to $-10.0$ quench is unable to achieve a completely relaxed state. 

It is known that a linear increase in entropy is the precursor of thermalization~\cite{Borgonovi:2016}. 
In Fig.~\ref{fig:control-frag-alpha-0.7} (b), we indeed observe that $S_n(t)$ exhibits a linear increase when the interaction strength is quenched to the same magnitude. 
Gradually weakening the quench, though, the linear increase starts to disappear and instead  an initial slower ramp starts to develop. 
For $g_d=+100.0$ to $g_d=-10.0$ quench, the growth of $S_n(t)$ clearly exhibits three different time scales rather than a linear increase of the entropy. 
In the first interval (I, green background), there is a clear quadratic increase which ends up in a very short plateau.
In the second interval (II, yellow background), $S_n(t)$ increases more rapidly, acquiring the familiar linear trend of the stronger quenches.
In the third interval (III, red background), the growth of $S_n(t)$ slows down and does not quite reach the GOE saturation value.
Our findings are in perfect agreement with what observed in works dealing with two-body random interaction models described by superpositions of mean-field states that are used to describe the onset of chaotic dynamics~\cite{Flambaum:2001}. 
In these models, the predicted entropy dynamics follows an initial quadratic growth, followed by a linear growth, before reaching eventual saturation.  
The quadratic entropy growth is the more generic case in systems where the spreading function follows a Breit-Wigner form (Lorentzian). 
This Lorentzian shape arises when the interaction strength is such that the spreading width $\Gamma$ is small compared to the energy range over which the states are spread. 
The linear regime, instead, appears when the interactions are strong enough to cause the spreading function (a.k.a. the local density of states) to take a Gaussian form because many more states become populated (as it is the case in Fig.~\ref{fig:coefficients}). 
The linear growth of entropy in this regime reflects the system's chaotic behavior, where the many-body state explores the energy space more uniformly.

The slower entropy increase and the different functional behavior than the immediate linear increase followed by saturation encountered in the strong quenches suggests that the entropy evolution can generally be manipulated by playing around with the strength and reach of the interactions. 
This finding is in agreement with the notion that the presence of long-range forces, such as in gravitational or Coulombic systems, tends to slow down the relaxation towards equilibrium due to the non-local nature of the interactions~\cite{Dauxois:2002, Bouchet:2005, Levin:2014, Hauke:2013, Schachenmayer:2013}.
Thus, we conjecture that the entropy might be even slowed down entirely into a prethermal plateau following a similar interplay of interaction range and sudden quenches in related systems.

%%%%%%%%%%%%%%%%%%%%%%%%%%%%%%%%%%%%%%%%
%%%%%%%%%%%%%%%%%%%%%%%%%%%%%%%%%%%%%%%%

The dynamics shown in Fig.~\ref{fig:control-frag-alpha-2.7} for shorter range interactions and smaller quench is distinctly different. 
The system remains mostly insensitive to the reduction in the quench strength.
Some modulations in the amplitude are observed, but the basic conclusion remains unchanged: the system saturates to the GOE value in a single time scale, irrespective of the post-quench parameters. 
Fig~\ref{fig:control-frag-alpha-2.7}(b) shows that $S_n(t)$ maintains its linearity at all times which guarantees thermal-like (i.e. linear) relaxation for all weaker quench processes. 
No appreciable dynamics slowdown is observed.
Notably, the saturation value is almost independent of the postquench parameter. 
Even in this case, though, we can relate our findings to Ref.~\cite{Flambaum:2001}.
In the stronger quenches, there is a lot of energy injected into the system that make the particles interact very strongly from the get-go. 
As a result, only the linear regime and the eventual saturation can be detected.
The quadratic regime should still exist, but it is pushed to very small times that do not appear very visible over longer time scales.

We thus conclude that when $\alpha$ is smaller than the dimension of the system, the dynamics can be drastically modified by the strength of the quench protocol, whereas for shorter range interactions the dynamics remains rather insensitive to it.

%%%%%%%%%%%%%%%%%%%%%%%%%%%%%%%%%%%%%%%%
%%%%%%%%%%%%%%%%%%%%%%%%%%%%%%%%%%%%%%%%

\section{Conclusions and Outlook}
Understanding non-equilibrium dynamics and relaxation of quantum many-body systems is one of the major unsolved problems in quantum mechanics. 
The scarcity of appropriate theoretical tools to probe the relaxation dynamics makes the scenario even more challenging. 
When a generic isolated quantum system relaxes, the process of relaxation is not always known. 
There may be partial relaxation, or different stages of relaxation that can happen on different time scales leading to very complex nonequilibrium dynamics.

In this regard, thermalization is one of the most intriguing features and has been shown to emerge in various theoretical and experimental setups, although its physical origin is still not properly understood. 
Near-integrable quantum systems, which fail to thermalize on experimental time scales, can relax to some prethermal states described by the Gibbs ensemble. 
Existing numerical studies of the quench dynamics in the one-dimensional transverse field Ising model with long-range interactions $(\frac{1}{r^{\alpha}})$ exhibit signatures of prethermalization with subsequent dynamical phase transitions when $\alpha < d$ ($d$ being the dimension of the system). 

The present work addresses the nonequilibrium quench dynamics in a long-range interacting system.
We investigated a generalized long-range interacting Bose gas where the strong repulsive interactions are suddenly quenched to strong attractive values, following the seminal observation of the exotic sTG gas in Haller's experiment~\cite{Haller:2009}.
We presented a numerical study of the many-body dynamics for a wide range of interactions for both bosonic and fermionic systems. 
The usage of {\it ab initio} many-body methods facilitates a complete understanding of the exotic quenched phase in terms of fragmentation, delocalization and entropy production. 
We observed that the dynamics exhibited by the quench reveals new insights into the structure of quantum state relaxation. 

Although the system eventually relaxes to a maximal entropy state, it passes through very complex phases of violent fragmentation and chaotic delocalization. 
The entropy production is determined by the number of configurations in the Hilbert space, which are saturated for all the numbers of orbitals that can be employed with current computational capabilities.
The observation of unbounded entropy and chaotic spreading of the quenched gas in the available Hilbert space strongly suggests a behavior akin to that of a classical gas. 
We presented different measures of entropy for our discussion. 
However, we chose the occupation entropy as an order parameter to produce a parameter-time diagram that illustrates the relaxation process as a function of interaction length, covering the entire range of $0.5 < \alpha < 4.0$. 
We found an exponential boundary separating the relaxing (quantum) from the relaxed (classical) phase. 
The same relaxation behavior is observed for spinless fermions, which establishes the universality of such chaotic dynamics in interacting quantum systems.
We also studied a quench protocol to weaker attractions and observed signals of dynamics slowdown and potential precursors to prethermalization for truly long-ranged interactions. 

The theoretical findings obtained in this work for a broad range of decay exponents can be tested across various experimental systems such as trapped ions, dipolar systems, Rydberg states, and quantum gases in cavities. 
Each of these systems serves as a prototypical platform to examine quench dynamics in long-range physics.
The typical scenarios involving confined bosons and fermions in the strongly interacting limit, as studied here, can also be implemented in various other spin models. 
These include the quantum Ising chain and the XYZ model, both of which allow complete control over the spin-spin interactions.
Among these examples, trapped ions uniquely enable the realization of long-range interactions by adjusting the decay exponent. 
This adjustment is achieved by tuning the frequency of the detuned laser beat note and the trap frequency.
Controlling the decay constant is fundamental to the ethos of quantum simulation. 
In the realm of quantum computation, trapped-ion systems are routinely examined, and qubit control is facilitated through the manipulation of long-range interactions.
Therefore, the comprehensive analysis of the quench protocol presented in this manuscript can be effectively used as a quantum tool to manipulate dynamics and to explore quantum information tasks within the highly versatile platform of trapped-ion systems.

\section*{Acknowledgements} \label{sec:acknowledgements}
We gratefully acknowledge computation time on the ETH Zurich Euler cluster and at the High-Performance Computing Center Stuttgart (HLRS). 
We thank E. Bergholtz for useful discussions. 
This work is partly supported by the Swedish Research Council (2018-00313) and Knut and Alice Wallenberg Foundation (KAW) via the project Dynamic Quantum Matter (2019.0068).

\appendix
\section{Appendix - Initial states} 
\label{sec:initial-states}

\begin{figure}[h!]
    \centering
    \includegraphics[width=\columnwidth]{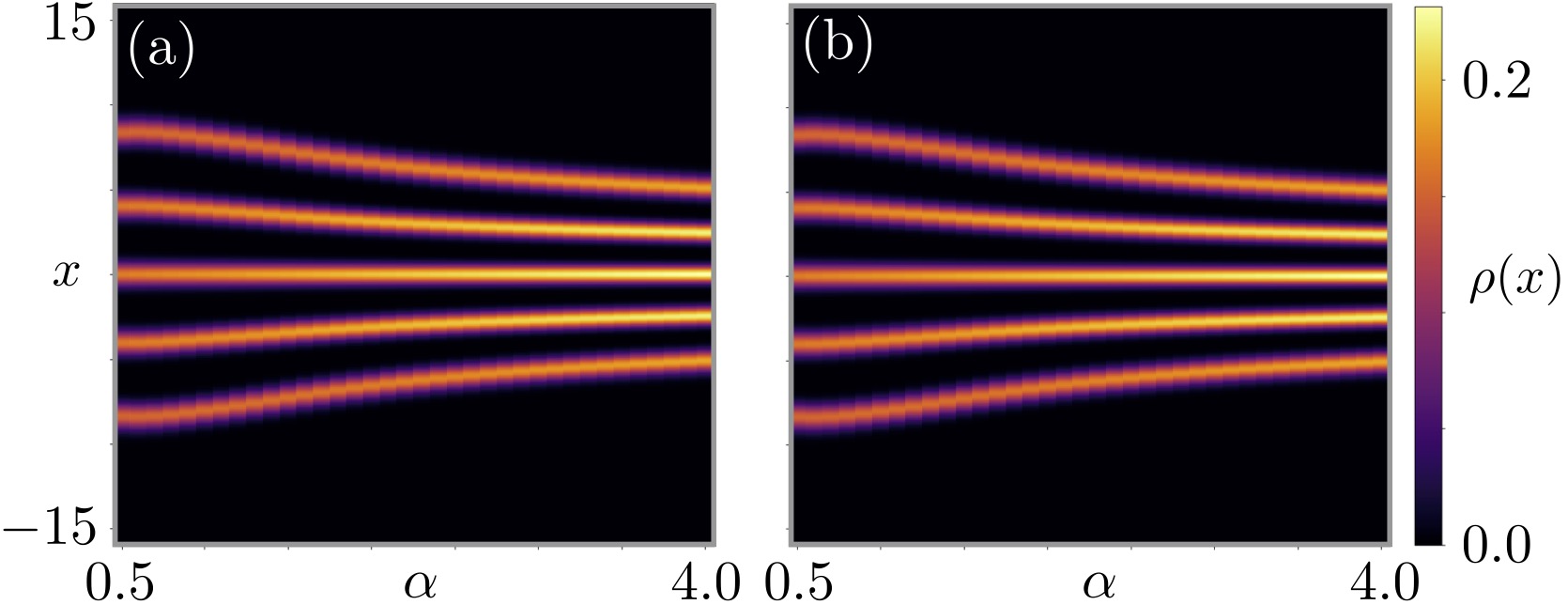}
    \caption{Real-space density distribution $\rho(x)$ as a function of the interaction power $alpha$ for the crystal states used as initial states in the quench procedures of the main text. (a) $N=5$ repulsive bosons, (b) $N=5$ repulsive fermions. For both panels an interaction strength $g_d=100.0$ and $M=10$ orbitals were used.}
    \label{fig:initial-states}
\end{figure}

In this section we show the density profile for the initial states used in the quench protocol. 
The initial state for $N=5$ bosons is presented in  Fig.~\ref{fig:initial-states} (a).
The initial state for $N=5$ fermions is presented in  Fig.~\ref{fig:initial-states} (b). 
The real space density distribution is plotted as a function of the exponent $\alpha$.
The interaction strength used is $g_d=100.0$ and the computation is performed with $M=10$ orbitals. 
We have verified that employing a higher number of orbitals does not change the density distribution.

\section{Appendix - Stronger quench dynamics} 
\label{sec:stronger-quench}
\begin{figure}
    \centering
    \includegraphics[width=0.7\columnwidth, angle=0]{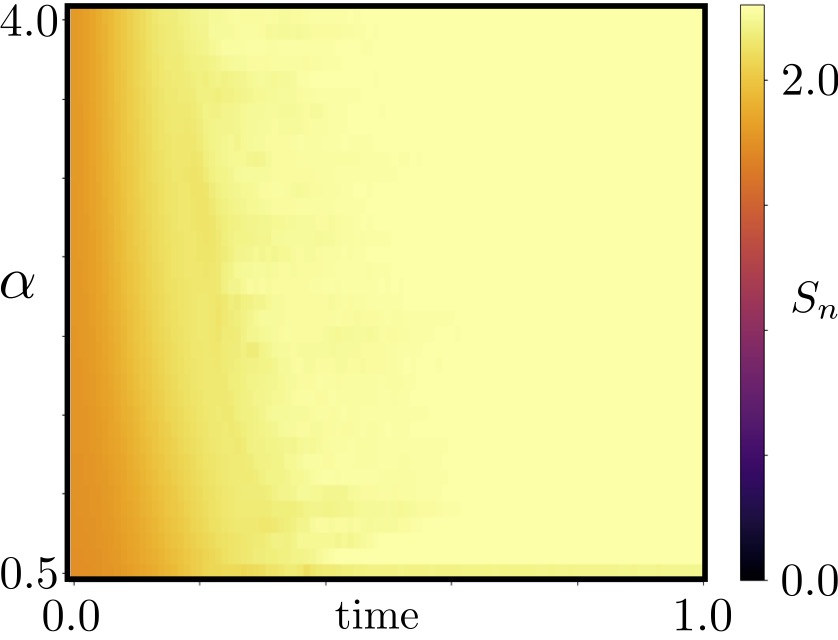}
    \caption{Occupation entropy dynamics as a function of interaction power $\alpha$ for $N=5$ quenched bosons with $g_d=100.0 \to -200.0$ (i.e. a stronger quench than in the main text) and $M=10$.}
    \label{fig:PD-bosons-gd--200}
\end{figure}

In this section,  we visualize and discuss the phase diagram of the relaxation process across different values of interaction exponent $\alpha$ for an even stronger quench than the one considered in the main text.
This is shown in Fig.~\ref{fig:PD-bosons-gd--200}.
Compared to the phase diagrams presented in the main text, we observe two different features.
First, the relaxation process occurs at even shorter times than for the quench $g_d =100.0 \to -100.0$, which is expected given the stronger interactions.
Second, the boundary between the relaxing and relaxed states loses its exponential character almost completely, and instead appears to be better described by a linear function. 

%%%%%%%%%%%%%%%%%%%%%%%%%%%%%%%%%%%%%%%%
%%%%%%%%%%%%%%%%%%%%%%%%%%%%%%%%%%%%%%%%

%%%%%%%%%%%%%%%%%%%%%%%%%%%%%%%%%%%%%%%%
%%%%%%%%%%%%%%%%%%%%%%%%%%%%%%%%%%%%%%%%

%%%%%%%%%%%%%%%%%%%%%%%%%%%%%%%%%%%%%%%%%%%%%%%%%%%%%%%%%%%%%%%%%%%%%%%%%%%%%%%%%%%%%%%%%%%%%
\section{Appendix - Units}

In this appendix, we discuss the units for the simulations presented in the main text.
The system consists of $N=5$ bosons or fermions in an optical harmonic trap of frequency $\omega$, leading to the external potential $\hat{V}(x) = \frac{1}{2} \omega x^2$.

The units of our simulations are chosen as follows.
We choose to set the unit of length as the inverse of the harmonic trapping frequency, i.e. $\bar{L} \equiv \sqrt{\hbar/(m\omega)}$.
We then run simulations with 1024 gridpoints in an interval $x \in [-16 \bar{L}, 16 \bar{L}]$, giving a resolution of 0.03125 $\bar{L}$.

The unit of energy $\bar{E}$ in MCTDH-X is defined in terms of the unit of length as $\bar{E} \equiv \frac{\hbar^2}{m \bar{L}^2}$. By inserting our choice for the unit of length, we immediately see that the unit of energy corresponds to the quantized energy of the harmonic trap, i.e. $\bar{E} = \hbar \omega$.
During the quench procedure, we change the strength of the long-range interactions from an initial value of $g_d=+100 \bar{E}$ to a negative (attractive) value of between $g_d=-10 \bar{E}$ and $g_d=-100 \bar{E}$, as explained in the main text.
Note that while strong, the initial repulsion is not enough for the particles to escape the harmonic confinement, as can be seen from Fig.~\ref{fig:initial-states}.

The unit of time in MCTDH-X is also defined from the unit of length, as 
$\bar{t} \equiv \frac{m \hat{L}^2}{\hbar}$.
Again inserting our definition of unit of length, we can simplify $\bar{t}=\frac{1}{\omega}$, i.e. the unit of time is the inverse frequency of the trap. 
In our simulations, we ran most time evolutions up until around $t \approx 1.2 \bar{t}$ as this time scale is enough to probe the relaxation dynamics of the quench.
Note that this is a relatively short time -- just slightly above the time it takes for one complete oscillation in a harmonic potential.
The fast dynamics we are generating is a consequence of the strong interactions employed.

\newpage
\bibliographystyle{unsrt}
\bibliography{biblio}

\begin{thebibliography}{100}

\bibitem{Defenu2021}
Nicolò Defenu, Tobias Donner, Tommaso Macrì, Guido Pagano, Stefano Ruffo, and
  Andrea Trombettoni.
\newblock Long-range interacting quantum systems.
\newblock {\em Rev. Mod. Phys.}, 95:035002, 2023.

\bibitem{Saffman}
M.~Saffman, T.~G. Walker, and K.~Mølmer.
\newblock Quantum information with rydberg atoms.
\newblock {\em Rev. Mod. Phys.}, 82:2313, 2010.

\bibitem{Monroe:2021}
C.~Monroe et. al.
\newblock Programmable quantum simulations of spin systems with trapped ions.
\newblock {\em Rev. Mod. Phys.}, 93:025001, 2021.

\bibitem{Schneider:2012}
Ch~Schneider, Diego Porras, and Tobias Schaetz.
\newblock Experimental quantum simulations of many-body physics with trapped
  ions.
\newblock {\em Rep. Prog. Phys.}, 75:024401, 2012.

\bibitem{Carr:2009}
Lincoln~D Carr, David DeMille, Roman~V Krems, and Jun Ye.
\newblock Cold and ultracold molecules: science, technology and applications.
\newblock {\em New J. Phys.}, 11:055049, 2009.

\bibitem{Lahaye:2009}
T.~Lahaye, C.~Menotti, L.~Santos, M.~Lewenstein, and T.~Pfau.
\newblock The physics of dipolar bosonic quantum gases.
\newblock {\em Rep. Prog. Phys.}, 72:126401, 2009.

\bibitem{Islam:2013}
R.~Islam, C.~Senko, W.~C. Campbell, S.~Korenblit, J.~Smith, A.~Lee, E.~E.
  Edwards, C.~C.~J. Wang, J.~K. Freericks, and C.~Monroe.
\newblock Emergence and frustration of magnetism with variable-range
  interactions in a quantum simulator.
\newblock {\em Science}, 340:583, 2013.

\bibitem{Jurcevic:2014}
P.~Jurcevic, B.~Lanyon, and P.~et~al. Hauke.
\newblock Quasiparticle engineering and entanglement propagation in a quantum
  many-body system.
\newblock {\em Nature}, 511:202, 2014.

\bibitem{Cirac}
D.~Porras and J.~I. Cirac.
\newblock Effective quantum spin systems with trapped ions.
\newblock {\em Phys. Rev. Lett.}, 92:207901, 2004.

\bibitem{Britton}
J.~Britton, B.~Sawyer, and A.~et~al. Keith.
\newblock Engineered two-dimensional ising interactions in a trapped-ion
  quantum simulator with hundreds of spins.
\newblock {\em Nature}, 484:489, 2012.

\bibitem{Islam}
R.~Islam, C.~Senko, W.~Campbell, S.~Korenblit, J.~Smith, A.~Lee, E.~Edwards,
  C.-C. Wang, J.~Freericks, and C.~Monroe.
\newblock Emergence and frustration of magnetism with variable-range
  interactions in a quantum simulator.
\newblock {\em Science}, 340:583, 2013.

\bibitem{Jens:2013}
Jens Eisert, Mauritz van~den Worm, Salvatore~R. Manmana, and Michael Kastner.
\newblock Breakdown of quasilocality in long-range quantum lattice models.
\newblock {\em Phys. Rev. Lett.}, 111:260401, 2013.

\bibitem{Schachenmayer:2013}
J.~Schachenmayer, B.~P. Lanyon, C.~F. Roos, and A.~J. Daley.
\newblock Entanglement growth in quench dynamics with variable range
  interactions.
\newblock {\em Phys. Rev. X}, 3:031015, 2013.

\bibitem{Santos:2016}
Lea~F. Santos, Fausto Borgonovi, and Giuseppe~Luca Celardo.
\newblock Cooperative shielding in many-body systems with long-range
  interaction.
\newblock {\em Phys. Rev. Lett.}, 116:250402, 2016.

\bibitem{Anton:2016}
Anton~S. Buyskikh, Maurizio Fagotti, Johannes Schachenmayer, Fabian Essler, and
  Andrew~J. Daley.
\newblock Entanglement growth and correlation spreading with variable-range
  interactions in spin and fermionic tunneling models.
\newblock {\em Phys. Rev. A}, 93:053620, 2016.

\bibitem{Stefan}
Stefan Schütz and Giovanna Morigi.
\newblock Prethermalization of atoms due to photon-mediated long-range
  interactions.
\newblock {\em Phys. Rev. Lett.}, 113:203002, 2016.

\bibitem{Michael}
Michael Kastner.
\newblock Diverging equilibration times in long-range quantum spin models.
\newblock {\em Phys. Rev. Lett.}, 106:130601, 2011.

\bibitem{Li:2017}
Jun-Ru Li, Jeongwon Lee, Wujie Huang, Sean Burchesky, Boris Shteynas, Furkan
  Çağrı Top, Alan~O. Jamison, and Wolfgang Ketterle.
\newblock A stripe phase with supersolid properties in spin–orbit-coupled
  bose–einstein condensates.
\newblock {\em Nature}, 543:91, 2017.

\bibitem{Boettcher:2019}
Fabian Böttcher, Jan-Niklas Schmidt, Matthias Wenzel, Jens Hertkorn, Mingyang
  Guo, Tim Langen, and Tilman Pfau.
\newblock Transient supersolid properties in an array of dipolar quantum
  droplets.
\newblock {\em Phys. Rev. X}, 9:011051, 2019.

\bibitem{Tanzi:2019}
L.~Tanzi, E.~Lucioni, F.~Famà, J.~Catani, A.~Fioretti, C.~Gabbanini, R.~N.
  Bisset, L.~Santos, and G.~Modugno.
\newblock Observation of a dipolar quantum gas with metastable supersolid
  properties.
\newblock {\em Phys. Rev. Lett.}, 122:130405, 2019.

\bibitem{Tanzi:2019-2}
L.~Tanzi, S.~M. Roccuzzo, E.~Lucioni, F.~Famà, A.~Fioretti, C.~Gabbanini,
  G.~Modugno, A.~Recati, and S.~Stringari.
\newblock Supersolid symmetry breaking from compressional oscillations in a
  dipolar quantum gas.
\newblock {\em Nature}, 574:382, 2019.

\bibitem{Chomaz:2019}
L.~Chomaz, D.~Petter, P.~Ilzhöfer, G.~Natale, A.~Trautmann, C.~Politi,
  G.~Durastante, R.~M.~W. van Bijnen, A.~Patscheider, M.~Sohmen, M.~J. Mark,
  and F.~Ferlaino.
\newblock Long-lived and transient supersolid behaviors in dipolar quantum
  gases.
\newblock {\em Phys. Rev. X}, 9:021012, 2019.

\bibitem{Natale:2019}
G.~Natale, R.~M.~W. van Bijnen, A.~Patscheider, D.~Petter, M.~J. Mark,
  L.~Chomaz, and F.~Ferlaino.
\newblock Excitation spectrum of a trapped dipolar supersolid and its
  experimental evidence.
\newblock {\em Phys. Rev. Lett.}, 123:050402, 2019.

\bibitem{Tanzi:2021}
L.~Tanzi, J.~G. Maloberti, G.~Biagioni, A.~Fioretti, C.~Gabbanini, and
  G.~Modugno.
\newblock Evidence of superfluidity in a dipolar supersolid from nonclassical
  rotational inertia.
\newblock {\em Science}, 371(6534):1162--1165, 2021.

\bibitem{Norcia:2021}
Matthew~A. Norcia, Claudia Politi, Lauritz Klaus, Elena Poli, Maximilian
  Sohmen, Manfred~J. Mark, Russell~N. Bisset, Luis Santos, and Francesca
  Ferlaino.
\newblock Two-dimensional supersolidity in a dipolar quantum gas.
\newblock {\em Nature}, 596:357, 2021.

\bibitem{Sohmen:2021}
Maximilian Sohmen, Claudia Politi, Lauritz Klaus, Lauriane Chomaz, Manfred~J.
  Mark, Matthew~A. Norcia, and Francesca Ferlaino.
\newblock Birth, life, and death of a dipolar supersolid.
\newblock {\em Phys. Rev. Lett.}, 126:233401, 2021.

\bibitem{Sanchez-Baena:2023}
J.~Sánchez-Baena, C.~Politi, F.~Maucher, F.~Ferlaino, and T.~Pohl.
\newblock Heating a dipolar quantum fluid into a solid.
\newblock {\em Nature Communications}, 14:1868, 2023.

\bibitem{Recati:2023}
A.~Recati and S.~Stringari.
\newblock Supersolidity in ultracold dipolar gases.
\newblock {\em Nature Review Physics}, 5:735, 2023.

\bibitem{Kessler:2021}
Hans Keßler, Phatthamon Kongkhambut, Christoph Georges, Ludwig Mathey,
  Jayson~G. Cosme, and Andreas Hemmerich.
\newblock Observation of a dissipative time crystal.
\newblock {\em Phys. Rev. Lett.}, 127:043602, 2021.

\bibitem{Kongkhambut:2022}
Phatthamon Kongkhambut, Jim Skulte, Ludwig Mathey, Jayson~G. Cosme, Andreas
  Hemmerich, and Hans Keßler.
\newblock Observation of a continuous time crystal.
\newblock {\em Science}, 377(6606):670--673, 2022.

\bibitem{Choi}
S.~Choi, J.~Choi, and R.~et~al. Landig.
\newblock Observation of discrete time-crystalline order in a disordered
  dipolar many-body system.
\newblock {\em Nature}, 543:221, 2017.

\bibitem{Zhang}
J.~Zhang, P.~Hess, and A.~et~al. Kyprianidis.
\newblock Observation of a discrete time crystal.
\newblock {\em Nature}, 543:217, 2017.

\bibitem{Li:2019}
Jiaming Li, Andrew~K. Harter, Ji~Liu, Leonardo de~Melo, Yogesh~N. Joglekar, and
  Le~Luo.
\newblock Observation of parity-time symmetry breaking transitions in a
  dissipative floquet system of ultracold atoms.
\newblock {\em Nature Communications}, 10:855, 2019.

\bibitem{Wintersperger:2020}
Karen Wintersperger, Christoph Braun, F.~Nur Ünal, André Eckardt, Marco~Di
  Liberto, Nathan Goldman, Immanuel Bloch, and Monika Aidelsburger.
\newblock Realization of an anomalous floquet topological system with ultracold
  atoms.
\newblock {\em Nature Physics}, 16:1058, 2020.

\bibitem{Bracamontes:2022}
C.~A. Bracamontes, J.~Maslek, and J.~V. Porto.
\newblock Realization of a floquet-engineered moat band for ultracold atoms.
\newblock {\em Phys. Rev. Lett.}, 128:213401, 2022.

\bibitem{Sun:2023}
Bo-Ye Sun, Nathan Goldman, Monika Aidelsburger, and Marin Bukov.
\newblock Engineering and probing non-abelian chiral spin liquids using
  periodically driven ultracold atoms.
\newblock {\em PRX Quantum}, 4:020329, 2023.

\bibitem{Zhang:2023}
Jin-Yi Zhang, Chang-Rui Yi, Long Zhang, Rui-Heng Jiao, Kai-Ye Shi, Huan Yuan,
  Wei Zhang, Xiong-Jun Liu, Shuai Chen, and Jian-Wei Pan.
\newblock Tuning anomalous floquet topological bands with ultracold atoms.
\newblock {\em Phys. Rev. Lett.}, 130:043201, 2023.

\bibitem{Eriko}
Eriko Kaminishi, Takashi Mori, Tatsuhiko~N. Ikeda, and Masahito Ueda.
\newblock Entanglement prethermalization in the tomonaga-luttinger model.
\newblock {\em Phys. Rev. Lett.}, 97:013622, 2018.

\bibitem{Takuya}
Takuya Kitagawa, Adilet Imambekov, Jörg Schmiedmayer, and Eugene Demler.
\newblock The dynamics and prethermalization of one-dimensional quantum systems
  probed through the full distributions of quantum noise.
\newblock {\em New J. Phys.}, 13:073018, 2011.

\bibitem{Kaminishi}
E.~Kaminishi, T.~Mori, T.~N. Ikeda, and M.~Ueda.
\newblock Entanglement pre-thermalization in a one-dimensional bose gas.
\newblock {\em Nature Phys}, 11:1050, 2015.

\bibitem{Langen:2016}
Tim Langen, Thomas Gasenzer, and Jörg Schmiedmayer.
\newblock Prethermalization and universal dynamics in near-integrable quantum
  systems.
\newblock {\em Journal of Statistical Mechanics: Theory and Experiment}, page
  064009, 2016.

\bibitem{Tang:2018}
Yijun Tang, Wil Kao, Kuan-Yu Li, Sangwon Seo, Krishnanand Mallayya, Marcos
  Rigol, Sarang Gopalakrishnan, and Benjamin~L. Lev.
\newblock Thermalization near integrability in a dipolar quantum newton’s
  cradle.
\newblock {\em Phys. Rev. X}, 8:021030, 2018.

\bibitem{Marcos:2019}
Krishnanand Mallayya, Marcos Rigol, and Wojciech~De Roeck.
\newblock Prethermalization and thermalization in isolated quantum systems.
\newblock {\em Phys. Rev. X}, 9:021027, 2019.

\bibitem{Worm}
Mauritz van~den Worm, Brian~C Sawyer, John~J Bollinger, and Michael Kastner.
\newblock Relaxation timescales and decay of correlations in a long-range
  interacting quantum simulator.
\newblock {\em New J. Phys.}, 15:083007, 2013.

\bibitem{Kastner}
M.~Kastner and M.~van~den Worm.
\newblock Relaxation timescales and prethermalization in d-dimensional
  long-range quantum spin models.
\newblock {\em Phys. Scr.}, page 014039, 2015.

\bibitem{Mori}
T.~Mori.
\newblock Prethermalization in the transverse-field ising chain with long-range
  interactions.
\newblock {\em J. Phys. A: Math. Theor.}, 52:054001, 2019.

\bibitem{Rigol}
M.~Rigol, V.~Dunjko, and M.~Olshanii.
\newblock Thermalization and its mechanism for generic isolated quantum
  systems.
\newblock {\em Nature}, 452:854, 2008.

\bibitem{Peter}
Peter Reimann and Michael Kastner.
\newblock Equilibration of isolated macroscopic quantum systems.
\newblock {\em New J. Phys.}, 14:043020, 2012.

\bibitem{Anthony}
Anthony~J Short and Terence~C Farrelly.
\newblock Quantum equilibration in finite time.
\newblock {\em New J. Phys.}, 14:013063, 2012.

\bibitem{Berges}
J.~Berges, Sz. Borsányi, and C.~Wetterich.
\newblock Prethermalization.
\newblock {\em Phys. Rev. Lett.}, 93:142002, 2004.

\bibitem{Mori:2018}
T.~Mori, T.~N. Ikeda, E.~Kaminishi, and M.~Ueda.
\newblock Thermalization and prethermalization in isolated quantum systems: a
  theoretical overview.
\newblock {\em J. Phys. B: At. Mol. Opt. Phys.}, 51:112001, 2018.

\bibitem{Gring:2012}
M.~Gring, M.~Kuhnert, T.~Langen, T.~Kitagawa, B.~Rauer, M.~Schreitl, I.~Mazets,
  D.~Adu Smith, E.~Demler, and J.~Schmiedmay.
\newblock Relaxation and prethermalization in an isolated quantum system.
\newblock {\em Science}, 337:1318, 2012.

\bibitem{Francisco}
Francisco Machado, Dominic~V. Else, Gregory~D. Kahanamoku-Meyer, Chetan Nayak,
  and Norman~Y. Yao.
\newblock Long-range prethermal phases of nonequilibrium matter.
\newblock {\em Phys. Rev. X}, 10:011043, 2020.

\bibitem{Gong:2013}
Zhe~Xuan Gong and L~M Duan.
\newblock Prethermalization and dynamic phase transition in an isolated trapped
  ion spin chain.
\newblock {\em New J. Phys.}, 15:113051, 2013.

\bibitem{Marcus}
Marcus Kollar and Martin Eckstein.
\newblock Relaxation of a one-dimensional mott insulator after an interaction
  quench.
\newblock {\em Phys. Rev. Lett.}, 78:013626, 2008.

\bibitem{Martin}
Martin Eckstein, Marcus Kollar, and Philipp Werner.
\newblock Relaxation of a one-dimensional mott insulator after an interaction
  quench.
\newblock {\em Phys. Rev. Lett.}, 103:056403, 2009.

\bibitem{Bruno}
Bruno Bertini, Fabian~H.L. Essler, Stefan Groha, and Neil~J. Robinson.
\newblock Prethermalization and thermalization in models with weak
  integrability breaking.
\newblock {\em Phys. Rev. Lett.}, 115:180601, 2015.

\bibitem{Chomaz:2023}
Lauriane Chomaz, Igor Ferrier-Barbut, Francesca Ferlaino, Bruno Laburthe-Tolra,
  Benjamin~L Lev, and Tilman Pfau.
\newblock Dipolar physics: a review of experiments with magnetic quantum gases.
\newblock {\em Rep. Prog. Phys.}, 86:026401, 2023.

\bibitem{Patscheider}
A.~et~al. Patscheider.
\newblock Controlling dipolar exchange interactions in a dense
  three-dimensional array of large-spin fermions.
\newblock {\em Phys. Rev. Res.}, 2:023050, 2020.

\bibitem{Kaufman:2023}
A.~M. Kaufman and K.-K. Ni.
\newblock Quantum science with optical tweezer arrays of ultracold atoms and
  molecules.
\newblock {\em Nat. Phys.}, 17:1324, 2021.

\bibitem{Li-JR:2023}
J.-R. Li and et~al.
\newblock Tunable itinerant spin dynamics with polar molecules.
\newblock {\em Nature}, 614:70, 2023.

\bibitem{Lieb}
Elliott~H. Lieb and Werner Liniger.
\newblock Exact analysis of an interacting bose gas. i. the general solution
  and the ground state.
\newblock {\em Phys. Rev.}, 130:1605, 1963.

\bibitem{Kinoshita:2004}
Toshiya Kinoshita, Trevor Wenger, and David~S. Weiss.
\newblock Observation of a one-dimensional tonks-girardeau gas.
\newblock {\em Science}, 305:1125, 2004.

\bibitem{Astrakharchik:2005}
G.~E. Astrakharchik, J.~Boronat, J.~Casulleras, and S.~Giorgini.
\newblock Beyond the tonks-girardeau gas: Strongly correlated regime in
  quasi-one-dimensional bose gases.
\newblock {\em Phys. Rev. Lett.}, 95:190407, 2005.

\bibitem{Astrakharchik:2008}
G.~E. Astrakharchik and Yu.~E. Lozovik.
\newblock Super-tonks-girardeau regime in trapped one-dimensional dipolar
  gases.
\newblock {\em Phys. Rev. A}, 77:013404, 2008.

\bibitem{Haller:2009}
Elmar Haller, Mattias Gustavsson, Manfred~J. Mark, Johann~G. Danzl, Russel
  Hart, Guido Pupillo, and Hanns-Christoph Nägerl.
\newblock Realization of an excited, strongly correlated quantum gas phase.
\newblock {\em Science}, 325:1224, 2009.

\bibitem{Muth:2010}
Dominik Muth and Michael Fleischhauer.
\newblock Dynamics of pair correlations in the attractive lieb-liniger gas.
\newblock {\em Phys. Rev. Lett.}, 105:150403, 2010.

\bibitem{Tschischik:2015}
W.~Tschischik and M.~Haque.
\newblock Repulsive-to-attractive interaction quenches of a one-dimensional
  bose gas in a harmonic trap.
\newblock {\em Phys. Rev. A}, 91:053607, 2015.

\bibitem{Chen:2010}
Shu Chen, Liming Guan, Xiangguo Yin, Yajiang Hao, and Xi-Wen Guan.
\newblock Transition from a tonks-girardeau gas to a super-tonks-girardeau gas
  as an exact many-body dynamics problem.
\newblock {\em Phys. Rev. A}, 81:031609(R), 2010.

\bibitem{Streltsov:2006}
Alexej~I. Streltsov, Ofir~E. Alon, and Lorenz~S. Cederbaum.
\newblock General variational many-body theory with complete self-consistency
  for trapped bosonic systems.
\newblock {\em Phys. Rev. A}, 73:063626, 2006.

\bibitem{Streltsov:2007}
Alexej~I. Streltsov, Ofir~E. Alon, and Lorenz~S. Cederbaum.
\newblock Role of excited states in the splitting of a trapped interacting
  bose-einstein condensate by a time-dependent barrier.
\newblock {\em Phys. Rev. Lett.}, 99:030402, 2007.

\bibitem{Alon:2007}
Ofir~E. Alon, Alexej~I. Streltsov, and Lorenz~S. Cederbaum.
\newblock Unified view on multiconfigurational time propagation for systems
  consisting of identical particles.
\newblock {\em J. Chem. Phys.}, 127:154103, 2007.

\bibitem{Alon:2008}
O.~E. Alon, A.~I. Streltsov, and L.~S. Cederbaum.
\newblock Multiconfigurational time-dependent hartree method for bosons:
  Many-body dynamics of bosonic systems.
\newblock {\em Phys. Rev. A}, 77:033613, 2008.

\bibitem{Lin:2023}
L.~Su, A.~Douglas, M.~Szurek, and et~al.
\newblock Dipolar quantum solids emerging in a hubbard quantum simulator.
\newblock {\em Nature}, 622:724, 2023.

\bibitem{Lode:2016}
A.~U.~J. Lode.
\newblock Multiconfigurational time-dependent hartree method for bosons with
  internal degrees of freedom: Theory and composite fragmentation of
  multicomponent bose-einstein condensates.
\newblock {\em Phys. Rev. A}, 93:063601.

\bibitem{Fasshauer:2016}
Elke Fasshauer and A.~U.~J. Lode.
\newblock Multiconfigurational time-dependent hartree method for fermions:
  Implementation, exactness, and few-fermion tunneling to open space.
\newblock {\em Phys. Rev. A}, 93:033635, 2016.

\bibitem{Lin:2020}
R.~Lin, P.~Molignini, L.~Papariello, M.~C. Tsatsos, C.~Lévêque, S.~E. Weiner,
  E.~Fasshauer, and R.~Chitra.
\newblock Mctdh-x: The multiconfigurational time-dependent hartree method for
  indistinguishable particles software.
\newblock {\em Quantum Sci. Technol.}, 5:024004, 2020.

\bibitem{Lode:2020}
A.~U.~J. Lode, C.~Lévêque, L.~B. Madsen, A.~I. Streltsov, and O.~E. Alon.
\newblock Colloquium: Multiconfigurational time-dependent hartree approaches
  for indistinguishable particles.
\newblock {\em Rev. Mod. Phys}, 92:011001, 2020.

\bibitem{MCTDHX}
A.~U.~J. Lode, M.~C. Tsatsos, E.~Fasshauer, S.~E. Weiner, R.~Lin,
  L.~Papariello, P.~Molignini, C.~Lévêque, M.~Büttner, J.~Xiang, S.~Dutta,
  and Y.~Bilinskaya.
\newblock Mctdh-x: The multiconfigurational time-dependent hartree method for
  indistinguishable particles software, 2024.

\bibitem{ultracold}
A.~U.~J. Lode, M.~C. Tsatsos, E.~Fasshauer amd S.~E.~Weiner, R.~Lin,
  L.~Papariello, P.~Molignini, C.~Lévêque, M.~Buettner, J.~Xiang, S.~Dutta,
  R.~Roy, Y.~Bilinskaya, and M.~Eder.
\newblock Mctdh-x: The multiconfigurational time-dependent hartree method for
  indistinguishable particles software.
\newblock 2024.

\bibitem{Lode:2012}
Axel U.~J. Lode, Kaspar Sakmann, Ofir~E. Alon, Lorenz~S. Cederbaum, and
  Alexej~I. Streltsov.
\newblock Numerically exact quantum dynamics of bosons with time-dependent
  interactions of harmonic type.
\newblock {\em Phys. Rev. A}, 86:063606, 2012.

\bibitem{Cao:2013}
L.~Cao, S.~Krönke, O.~Vendrell, and P.~Schmelcher.
\newblock The multi-layer multi-configuration time-dependent hartree method for
  bosons: Theory, implementation, and applications.
\newblock {\em J. Chem. Phys.}, 139:134103, 2013.

\bibitem{barnali_axel}
Axel U.~J. Lode, Barnali Chakrabarti, and Venkata K.~B. Kota.
\newblock Many-body entropies, correlations, and emergence of statistical
  relaxation in interaction quench dynamics of ultracold bosons.
\newblock {\em Phys. Rev. A}, 92:033622, 2015.

\bibitem{Fischer:2015}
Uwe~R. Fischer, Axel U.~J. Lode, and Budhaditya Chatterjee.
\newblock Condensate fragmentation as a sensitive measure of the quantum
  many-body behavior of bosons with long-range interactions.
\newblock {\em Phys. Rev. A}, 91:063621, 2015.

\bibitem{chatterjee:2018}
Budhaditya Chatterjee and Axel U.~J. Lode.
\newblock Order parameter and detection for a finite ensemble of crystallized
  one-dimensional dipolar bosons in optical lattices.
\newblock {\em Phys. Rev. A}, 98:053624, Nov 2018.

\bibitem{Molignini:2018}
Paolo Molignini, Luca Papariello, Axel U.~J. Lode, and R.~Chitra.
\newblock Superlattice switching from parametric instabilities in a
  driven-dissipative bose-einstein condensate in a cavity.
\newblock {\em Phys. Rev. A}, 98:053620, 2018.

\bibitem{Bera:2019}
S.~Bera, B.~Chakrabarti, A.~Gammal, M.~C. Tsatsos, M.~L. Lekala, B.~Chatterjee,
  C.~Lévêque, and A.~U.~J. Lode.
\newblock Sorting fermionization from crystallization in many-boson
  wavefunctions.
\newblock {\em Scientific Reports}, 9:17873, 2019.

\bibitem{Lin:2019}
Rui Lin, Luca Papariello, Paolo Molignini, R.~Chitra, and Axel U.~J. Lode.
\newblock Superfluid--mott-insulator transition of ultracold superradiant
  bosons in a cavity.
\newblock {\em Phys. Rev. A}, 100:013611, 2019.

\bibitem{chatterjee:2019}
Budhaditya Chatterjee, Marios~C Tsatsos, and Axel U~J Lode.
\newblock Correlations of strongly interacting one-dimensional ultracold
  dipolar few-boson systems in optical lattices.
\newblock {\em New Journal of Physics}, 21(3):033030, mar 2019.

\bibitem{chatterjee:2020}
Budhaditya Chatterjee, Camille Lévêque, Jörg Schmiedmayer, and Axel U.~J.
  Lode.
\newblock Detecting one-dimensional dipolar bosonic crystal orders via full
  distribution functions.
\newblock {\em Phys. Rev. Lett.}, 125:093602, Aug 2020.

\bibitem{Lin:2020-PRA}
Rui Lin, Paolo Molignini, Axel U.~J. Lode, and R.~Chitra.
\newblock Pathway to chaos through hierarchical superfluidity in blue-detuned
  cavity-bec systems.
\newblock {\em Phys. Rev. A}, 101:061602(R), 2020.

\bibitem{Lin:2021}
Rui Lin, Christoph Georges, Jens Klinder, Paolo Molignini, Miriam Büttner,
  Axel U.~J. Lode, R.~Chitra, Andreas Hemmerich, and Hans Kessler.
\newblock Mott transition in a cavity-boson system: A quantitative comparison
  between theory and experiment.
\newblock {\em SciPost Phys.}, 11:030, 2021.

\bibitem{Molignini:2022}
Paolo Molignini, Camille Lévêque, Hans Keßler, Dieter Jaksch, R.~Chitra, and
  Axel U.~J. Lode.
\newblock Crystallization via cavity-assisted infinite-range interactions.
\newblock {\em Phys. Rev. A}, 106:L011701, 2022.

\bibitem{Rosa-Medina:2022}
Rodrigo Rosa-Medina, Francesco Ferri, Fabian Finger, Nishant Dogra, Katrin
  Kroeger, Rui Lin, R.~Chitra, Tobias Donner, and Tilman Esslinger.
\newblock Observing dynamical currents in a non-hermitian momentum lattice.
\newblock {\em Phys. Rev. Lett.}, 128:143602, 2022.

\bibitem{Hughes:2023}
Michael Hughes, Axel U.~J. Lode, Dieter Jaksch, and Paolo Molignini.
\newblock Accuracy of quantum simulators with ultracold dipolar molecules: A
  quantitative comparison between continuum and lattice descriptions.
\newblock {\em Phys. Rev. A}, 107:033323, 2023.

\bibitem{Molignini:2024}
P.~Molignini and B.~Chakrabarti.
\newblock Super-tonks-girardeau quench of dipolar bosons in a one-dimensional
  optical lattice.
\newblock {\em arXiv:2401.10317}, 2024.

\bibitem{Bilinskaya:2024}
Y.~Bilinskaya, M.~Hughes, and P.~Molignini.
\newblock Exploring limits of dipolar quantum simulators with ultracold
  molecules.
\newblock {\em arXiv:2402.14914}, 2024.

\bibitem{Molignini:2024-2}
P.~Molignini.
\newblock Stability of quasicrystalline ultracold fermions to dipolar
  interactions.
\newblock {\em arXiv:2403.04830}, 2024.

\bibitem{Berman:2004}
G.P.Berman, F.~Borgonovi, and F.~M. Izrailev.
\newblock Irregular dynamics in a one-dimensional bose system.
\newblock {\em Phys. Rev. Lett.}, 92:030404, 2004.

\bibitem{Budhaditya}
Budhaditya Chatterjee and Axel U.~J. Lode.
\newblock Order parameter and detection for a finite ensemble of crystallized
  one-dimensional dipolar bosons in optical lattices.
\newblock {\em Phys. Rev. A}, 98:053624, 2018.

\bibitem{Santos:2012}
L.~F. Santos, F.~Borgonovi, and F.~M. Izrailev.
\newblock Chaos and statistical relaxation in quantum systems of interacting
  particles.
\newblock {\em Phys. Rev. Lett.}, 108:094102, 2012.

\bibitem{Kota}
V.~K.~B. Kota and R.~Sahu.
\newblock Single-particle entropy in (1+2)-body random matrix ensembles.
\newblock {\em Phys. Rev. E}, 66:037103, 2002.

\bibitem{Izrailev:2012}
L.~F. Santos, F.~Borgonovi, and 036209 F.~M. Izrailev Phys. Rev. E~85.
\newblock Onset of chaos and relaxation in isolated systems of interacting
  spins: Energy shell approach.
\newblock {\em Phys. Rev. E}, 85:036209, 2012.

\bibitem{Mark:1994}
Mark Srednicki.
\newblock Chaos and statistical relaxation in quantum systems of interacting
  particles.
\newblock {\em Phys. Rev. E}, 50:888, 1994.

\bibitem{Langen:2015}
T.~Langen, R.~Geiger, and H.-J. Schmiedmayer.
\newblock Ultracold atoms out of equilibrium.
\newblock {\em Annual Review of Condensed Matter Physics.}, 6:201, 2015.

\bibitem{Borgonovi:2016}
F.~Borgonovi, F.M. Izrailev, L.F. Santos, and V.G. Zelevinsky.
\newblock Quantum chaos and thermalization in isolated systems of interacting
  particles.
\newblock {\em Physics Reports}, 626:1, 2016.

\bibitem{Flambaum:2001}
V.~V. Flambaum1 and F.~M. Izrailev.
\newblock Dynamics and thermodynamics of systems with long-range interactions.
\newblock {\em Phys. Rev. E}, 64:036220, 2001.

\bibitem{Dauxois:2002}
T.~Dauxois, S.~Ruffo, E.~Arimondo, and M.~Wilkens.
\newblock Dynamics and thermodynamics of systems with long-range interactions.
\newblock {\em Lecture Notes in Physics}, 602, 2002.

\bibitem{Bouchet:2005}
F.~Bouchet, S.~Gupta, and D.~Mukamel.
\newblock Thermodynamics and dynamics of systems with long-range interactions.
\newblock {\em Physica A: Stat. Mech. Appl.}, 358:411, 2005.

\bibitem{Levin:2014}
Y.~Levin, Renato Pakter, Felipe~B. Rizzato, Tarcísio~N. Teles, and
  Fernanda~P.C. Benetti.
\newblock Nonequilibrium statistical mechanics of systems with long-range
  interactions.
\newblock {\em Phys. Rep.}, 535:1, 2014.

\bibitem{Hauke:2013}
P.~Hauke and L.~Tagliacozzo.
\newblock Spread of correlations in long-range interacting quantum systems.
\newblock {\em Phys. Rev. Lett.}, 111:207202, 2013.

\end{thebibliography}

\end{document}